\documentclass[aps,pra,twocolumn,superscriptaddress,floatfix]{revtex4-1}

\usepackage{graphicx}
\usepackage[usenames]{color}
\usepackage{dcolumn}
\usepackage{svn-multi}
\svnid{$Id: 120141219_LBCO_bozin 9063 2014-12-19 13:10:00Z bozin $}

\graphicspath{{./}}

\def\nba#1{}

\def\comment#1{}
\def\delete#1{}
\def\cuosix{CuO$_{6}$}
\def\cuotwo{CuO$_{2}$}

\def\lnsco{La$_{1.6-x}$Nd$_{0.4}$Sr$_{x}$CuO$_{4}$}

\def\ybco{YBa$_{2}$Cu$_{3}$O$_{6+d}$}
\def\lco{La$_{2}$CuO$_4$}
\def\lmo{LaMnO$_3$}
\def\lbco{La$_{2-x}$Ba$_x$CuO$_4$}
\def\lsco{La$_{2-x}$Sr$_x$CuO$_4$}

\def\rangexdoped{$0.095 \le x \le 0.155$}
\def\ranget{$15~{\rm K} \le T \le 550~{\rm K}$}
\def\bmab{Bmab}
\def\i4mmm{I4/mmm}
\def\f4mmm{F4/mmm}
\def\p42ncm{P4$_{2}$/ncm}
\def\pccn{Pccn}
\begin{document}
%

%
\title{Reconciliation of local and long range tilt correlations in underdoped La$\mathbf{_{2-x}}$Ba$\mathbf{_x}$CuO$\mathbf{_4}$ ~($\mathbf{0 \le x \le 0.155}$)}
%

%
\author{Emil~S. Bozin}
\affiliation{Department of Condensed Matter Physics and Materials Science, Brookhaven National Laboratory, Upton, New York 11973, USA}

\author{Ruidan Zhong}
\affiliation{Department of Condensed Matter Physics and Materials Science, Brookhaven National Laboratory, Upton, New York 11973, USA}
\affiliation{Materials Science and Engineering Department, Stony Brook University, Stony Brook, New York 11790, USA}

\author{Kevin~R. Knox}
\affiliation{Department of Condensed Matter Physics and Materials Science, Brookhaven National Laboratory, Upton, New York 11973, USA}

\author{Genda~Gu}
\affiliation{Department of Condensed Matter Physics and Materials Science, Brookhaven National Laboratory, Upton, New York 11973, USA}

\author{John~P. Hill}
\affiliation{Department of Condensed Matter Physics and Materials Science, Brookhaven National Laboratory, Upton, New York 11973, USA}

\author{John~M. Tranquada}
\affiliation{Department of Condensed Matter Physics and Materials Science, Brookhaven National Laboratory, Upton, New York 11973, USA}

\author{Simon~J.~L. Billinge}
\affiliation{Department of Condensed Matter Physics and Materials Science, Brookhaven National Laboratory, Upton, New York 11973, USA}
\affiliation{Department of Applied Physics and Applied Mathematics, Columbia University, New York, New York 10027, USA}
%

%
\date{\today}
%

%
\begin{abstract}
A long standing puzzle regarding the disparity of local and long range \cuosix\ octahedral tilt correlations in the underdoped regime of \lbco\ is addressed by utilizing complementary neutron powder diffraction and inelastic neutron scattering (INS) approaches. Long-range and static \cuosix\ tilt order with orthogonally inequivalent Cu-O bonds in the \cuotwo\ planes in the low temperature tetragonal (LTT) phase is succeeded on warming through the low-temperature transition by one with orthogonally equivalent bonds in the low temperature orthorhombic (LTO) phase. In contrast, the signatures of LTT-type tilts in the instantaneous local atomic structure persist on heating throughout the LTO crystallographic phase on the nanoscale, although becoming weaker as temperature increases. Analysis of the INS spectra for the $x=1/8$ composition reveals the dynamic nature of the LTT-like tilt fluctuations within the LTO phase and their 3D character. The doping dependence of relevant structural parameters indicates that the magnitude of the Cu-O bond anisotropy has a maximum at $x=1/8$ doping where bulk superconductivity is most strongly suppressed, suggesting that the structural anisotropy might be influenced by electron-phonon coupling and the particular stability of the stripe-ordered phase at this composition. The bond-length modulation that pins stripe order is found to be remarkably subtle, with no anomalous bond length disorder at low temperature, placing an upper limit on any in-plane Cu-O bondlength anisotropy. The results further reveal that although appreciable octahedral tilts persist through the high-temperature transition and into the high temperature tetragonal (HTT) phase, there is no significant preference between different tilt directions in the HTT regime. This study also exemplifies the importance of a systematic approach using complementary techniques when investigating systems exhibiting a large degree of complexity and subtle structural responses.
\end{abstract}
%


\maketitle

%
\section{Introduction}
\label{intro}

Since the discovery of high-temperature superconductivity in \lbco\ \cite{bedn86}, the connection between the lattice and electronic structures has been the subject of considerable attention \cite{jorg87,paul87}.   The further discovery \cite{mood88} of the ``1/8 anomaly''---a dramatic dip in the superconducting transition temperature $T_c$ at the dopant concentration of $x=1/8$---motivated careful powder diffraction studies of the structural phase diagram \cite{axe89,suzu89a,cox89,bill93}.  The latter revealed a low temperature transition ($\sim60$~K) to a crystal structure with inequivalent Cu-O bonds in orthogonal directions within the CuO$_2$ planes.  It is this structural anisotropy, associated with a particular pattern of tilts of the CuO$_6$ octahedra, that pins charge stripes \cite{tran95a,fuji04}.  While the charge stripes are compatible with superconducting correlations within the CuO$_2$ planes \cite{li07,tran08}, the $90^\circ$ rotation of the structural anisotropy from one layer to the next leads to frustration of the interlayer Josephson coupling and the depression of the onset of bulk superconducting order \cite{berg09b,hime02}.

Given the strong response of the electronic properties to the crystal symmetry, it is of considerable interest to understand the nature of the structural transitions in \lbco.  Both powder \cite{suzu89b,bill93} and single-crystal \cite{zhao07,huck11,wen12a} diffraction studies clearly demonstrate that the transition to the low-temperature structure, involving a change in tilt direction of the CuO$_6$ octahedra, is of first order.  A surprisingly different perspective is given by a pair-distribution-function (PDF) analysis of neutron scattering measurements \cite{bill94}.  In the latter analysis, the local structure appears to be unaffected by warming through the transition, retaining the low-temperature tilt pattern to higher temperatures.  That picture is supported by an x-ray absorption fine structure (XAFS) study \cite{hask00}.  Furthermore, the local tilts appear to remain in the high-temperature phase where the average tilts are zero \cite{hask00,bozi97,fabb13}.   The aim of the present paper is to resolve the apparent conflict between local and long-range measures of tilt correlations in \lbco.

\begin{figure}[tbf]
\includegraphics[width=80mm]{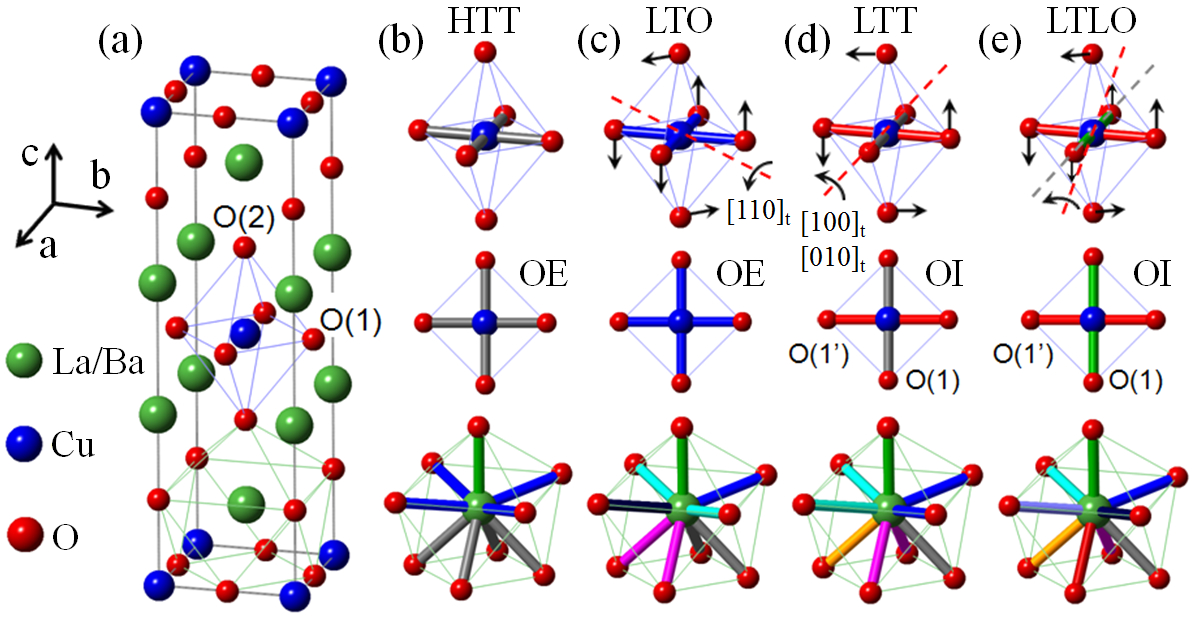}
\caption{\label{fig:LBCO-structuralDetails} (Color online) Structural details of \lbco. La/Ba are shown as large green spheres, Cu as intermediate blue spheres, and O as small red spheres. (a) Basic structural motif is shown for \i4mmm (HTT) phase, featuring CuO$_{6}$ octahedral unit and La/BaO$_{9}$ cage. O(1) and O(2) denote planar and apical oxygen respectively. Panels (b)-(e) highlight various aspects of the average crystal structures as follows: CuO$_{6}$ octahedral tilt symmetry (top row), in-plane bond-length distribution within CuO$_{4}$ plaquette (middle row), and dispersion of La/Ba-O distances within the La/BaO$_{9}$ cage (bottom row). Equal interatomic distances are represented by the same color. HTT denotes high temperature tetragonal ($I4/mmm$), LTO is low temperature orthorhombic ($Bmab$), LTT is low temperature tetragonal ($P4_2/ncm$), and LTLO is low temperature less orthorhombic ($Pccn$). The underlying in-plane symmetry is OE in HTT and LTO models, and OI in LTT and LTLO models, as indicated.}
\end{figure}

In terms of the symmetry of the average structure, the thermal sequence of structural transitions is well understood \cite{axe94}.  The high-temperature tetragonal (HTT) phase (space group $I4/mmm$) has the highest symmetry, with no octahedral tilts (see Fig.~\ref{fig:LBCO-structuralDetails}); here, the in-plane lattice parameters, $a_t=b_t$, correspond to the shortest Cu-Cu distance.  On cooling, there is a second-order transition to the low-temperature orthorhombic (LTO) phase (space group $Bmab$), involving tilts of the CuO$_6$ octahedra about [110]$_t$ axes of the HTT phase; as nearest-neighbor octahedra must rotate in opposite directions, the unit cell volume increases by a factor of two, with $a_o\approx b_o \approx \sqrt{2}a_t$.  (This is the structure of \lsco\ in the superconducting phase \cite{rada94}.)  Further cooling leads to a second transition involving a change in the octahedral tilt axis towards the $[100]_t$ and $[010]_t$ directions.  When the shift in the tilt axis is complete, the structure is the low-temperature tetragonal (LTT; space group $P4_2/ncm$), while a partial shift results in the low-temperature less-orthorhombic (LTLO; space group $Pccn$); in both cases, the unit cell size is the same as for LTO.

\begin{figure}[tbf]
\includegraphics[width=80mm]{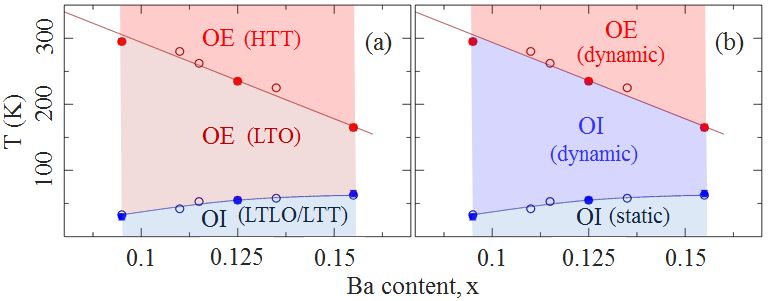}
\caption{\label{fig:AveragePhaseDiagram} (Color online) (a) Average structure ($x$, $T$) phase diagram of LBCO which has been replotted to highlight whether the structure is OE or OI. Open symbols and solid lines are from H{\" u}cker et al.~\cite{hucke;prb11}, solid symbols are from the present study. Shaded are regions of interest for this study having average OE (HTT and LTO) and OI (LTLO/LTT) symmetries, as indicated. (b) Local structure phase diagram with phase designations based on the results presented in this study.
Please see text for definitions of OE and OI.}
\end{figure}

For charge ordering, the key distinction among these different phases is the degree of symmetry among the Cu-O bonds.  The tilt pattern in the LTO phase leaves the Cu-O bonds in (approximately) orthogonal directions equivalent; we will denote such a symmetry as ``orthogonal equivalent'' (OE).  In contrast, the tilt of an octahedron about a $[100]_t$ axis, as in the LTT phase, leaves two in-plane oxygens within the CuO$_2$ plane but shifts the orthogonal pair above and below the plane, resulting in two different Cu-O bond lengths.  We will label this case as ``orthogonal inequivalent'' (OI), and note that the LTLO phase also has an OI symmetry.  The phase diagram for the relevant range of doping in \lbco\ is shown in Fig.~\ref{fig:AveragePhaseDiagram}(a).

The structural transformations have been analyzed in terms of a Landau-Ginzburg model, in which the degenerate octahedral tilts about $[110]_t$ and $[1\overline{1}0]_t$ axes are taken as the order parameters \cite{axe89,axe94}.  When only one order parameter is finite, one obtains the LTO phase; when both are finite and equal, the structure is LTT.  This model allows an elegant analysis of the phase diagram, and predicted the possible occurrence of the LTLO phase as confirmed in a closely-related system \cite{craw91}.   It has also motivated detailed studies of the octahedral tilts in \lsco\ \cite{birg87,thur89a,kimu00}, La$_{1.65}$Nd$_{0.35}$CuO$_4$ \cite{keim93}, and \lbco\ \cite{kimu05,waki06}; however, these studies have been done at points of reciprocal space that have a finite structure factor for both LTO and LTT-like tilts.  Hence, while these measurements have been interpreted in terms of LTO-like (OE) tilts, they could not uniquely distinguish the fluctuations from LTT-like (OI) tilts.

Helpful guidance is provided by a calculation of the potential-energy surface as a functional of octahedral tilts performed with density functional theory for a rough model of \lbco\ with $x=0.10$ \cite{pick91}.  The calculation finds that the lowest energy is given by LTT tilts, with local minima corresponding to LTO tilts higher in energy by about 15 meV.  It was proposed that the LTO phase might be stabilized with increasing temperature due to the entropy associated with low-energy octahedral tilt fluctuations \cite{pick91}.  This proposal gained support from a Monte Carlo study of the temperature dependence of a model including both the mean-field potential energy and interactions between neighboring octahedra \cite{cai94}.  The latter calculation yielded evidence for strong local LTT-like tilt amplitudes throughout the LTO phase.

In this paper, we present neutron total scattering measurements on polycrystalline samples of \lbco\ with $x=0.095$, 0.125, and 0.155 obtained as a function of temperature.  We analyze these data sets both by Rietveld refinement and by the pair distribution function analysis technique.  The two approaches yield complementary evidence for dynamical LTT-like tilts within the LTO phase, as well as local tilt fluctuations in the HTT phase.   We directly confirm the LTT-like tilt fluctuations in the LTO phase through inelastic neutron scattering measurements on a single crystal of \lbco\ with $x=0.125$.

The rest of the paper is organized as follows: In Sec.~\ref{exp} we describe the experimental and analysis methods, and the choice of reciprocal lattice used to index the reflections. In Sec.~\ref{results} we present three subsections dedicated to our results on average crystal structure, local structure, and octahedral tilt dynamics. In Sec.~\ref{discussion} we discuss the various properties as a function of the nominal Ba
content and temperature, compare our results with the literature, and in Sec.~\ref{summary} finish with a short summary.

%
\section{Experimental}
\label{exp}
%
Finely pulverized samples of \lbco , with Ba-content in the \rangexdoped\ range, as well as an undoped \lco\ polycrystalline reference, were grown using standard solid state protocols; these were used for the
total scattering atomic PDF experiments. Neutron time-of-flight measurements were carried out on the NPDF instrument at Los Alamos Neutron Scattering Center at
Los Alamos National Laboratory. Powders (15 grams each) were loaded under helium atmosphere into standard extruded vanadium containers and sealed. Temperature dependent
measurements in the \ranget\ range were performed using a closed cycle cryo-furnace sample environment for 2 hours at each temperature on each sample, yielding good statistics and a
favorable signal to noise ratio at high momentum transfers. Raw data were normalized and various experimental corrections performed following standard protocols~\cite{egami;b;utbp12}.
High resolution experimental PDFs were obtained from the Sine Fourier transform of the measured total scattering structure functions, $F(Q)$, over a broad range
of momentum transfers, $Q$ ($Q_{\rm max}  = 28$~\AA$^{-1}$). Data reduction to obtain the PDFs, $G(r)$, was carried out using the program {\sc PDFGETN}~\cite{peter;jac00}.
The average structure was assessed through Rietveld refinements~\cite{rietv;ac67} to the raw diffraction data using {\sc GSAS}~\cite{larso;unpub87} operated under
{\sc EXPGUI}~\cite{toby;jac01}, utilizing \i4mmm\ (HTT), \bmab\ (LTO), \p42ncm\ (LTT), and \pccn\ (LTLO) models from the literature~\cite{hucke;prb11}.
Structural refinement of PDF data was carried out using {\sc PDFFIT2} operated under {\sc PDFGUI}~\cite{farro;jpcm07} using the same models.

The single crystal of \lbco\ with $x=0.125$ was grown using the traveling-solvent floating zone method and has been characterized previously, as described elsewhere~\cite{huck11}.  The crystal, of size $\phi\sim8\,{\rm mm}\times20$~mm, was studied by inelastic neutron scattering using the HYSPEC instrument (beam line 14B) at the Spallation Neutron Source, Oak Ridge National Laboratory~\cite{winn14}.  For the experiment on HYSPEC, the crystal was mounted in a Displex closed-cycle cryostat.  With the $c$ axis vertical, scattering wave vectors ${\bf Q}=(H,K,0)$ are in the horizontal scattering plane. A fixed incident energy of 27 meV and a chopper frequency of 300 Hz were used for all data shown here, and the graphite-crystal array in the incident beam was put in the flat mode (no vertical focusing) to improve the resolution along $Q_z$.  For a typical measurement, the position-sensitive detector tank was placed at a particular mean scattering angle, and then measurements were collected for a series of sample orientations, involving rotations about the vertical axis in steps of $0.2^\circ$.  From such a set of scans, a four-dimensional data set was created and analyzed with the MANTID~\cite{arnold14} and DAVE~\cite{dave09} software packages.  Slices of data corresponding to particular planes in energy and wave-vector space can then be plotted from the larger data set.  Wave vectors will be expressed in units of $(2\pi/a,2\pi/b,2\pi/c)$ with $a=b=5.355$~\AA\ and $c=13.2$~\AA, corresponding to the LTT phase.

The measurements of the soft phonon that tilts along the Cu-O bonds were performed in the vicinity of the (330) position, which corresponds to a superlattice peak in the LTT but not the LTO phase.  To sample the fluctuations associated with the tilts of the LTO phase, it was necessary to tilt the sample so as to put (032) in the scattering plane.  We then looked at the behavior along $(H,3,2)$.

\begin{figure}[btf]
\includegraphics[width=80mm]{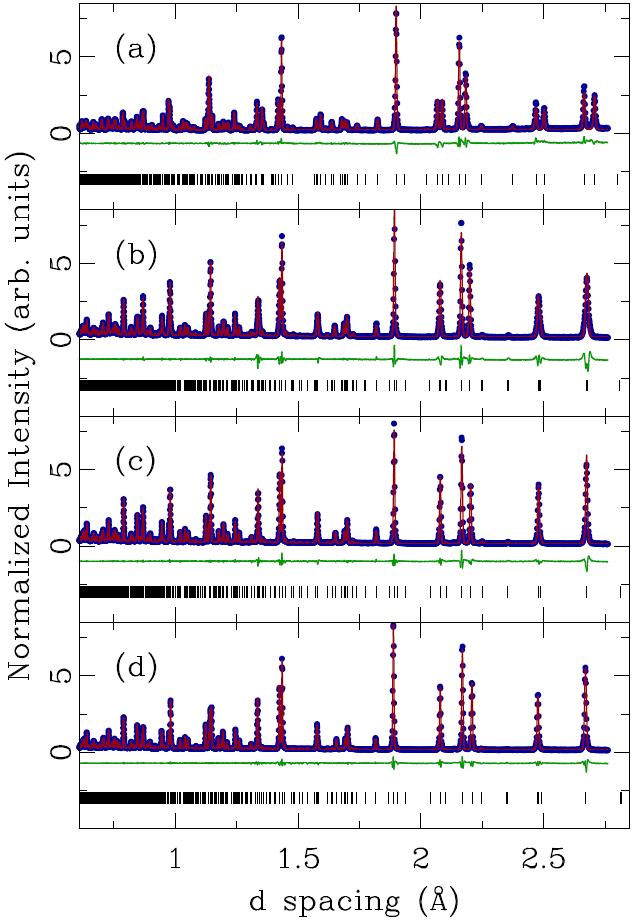}
\caption{\label{fig:RietveldFitsQuality} (Color online) Rietveld fits of the average structure models to LBCO data at 15~K. Closed blue symbols represent the data, solid red lines are the models, and solid green lines are the differences (offset for clarity). Vertical ticks mark reflections.  (a) x=0 using \bmab\ model, (b) x=0.095 using \pccn\ model, (c) x=0.125 using \p42ncm\ model, and (d) x=0.155 using \pccn\ model.}
\end{figure}

%
\section{Results}
\label{results}
%
%
\subsection{Average crystal structure}
\label{averagestructure}
%

\begin{figure}[tbf]
\includegraphics[width=80mm]{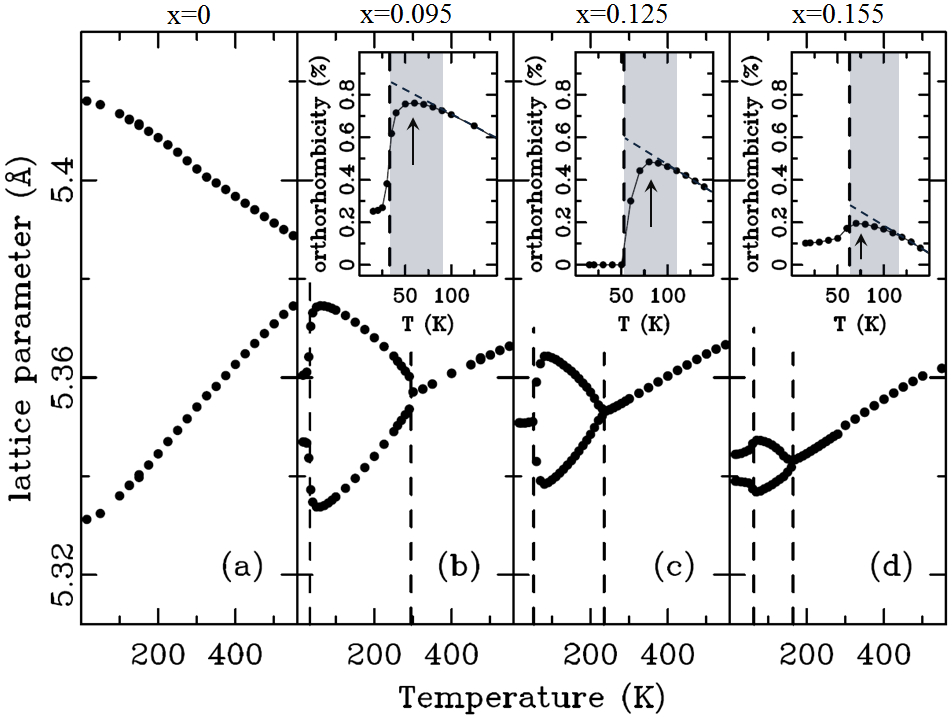}
\caption{\label{fig:lattice-fig} Temperature evolution of in-plane lattice parameters of \lbco\ obtained from Rietveld refinements for: x=0 (a), x=0.095 (b), x=0.125 (c), and x=0.155 (d). HTT parameters are given in \f4mmm setting. Vertical dashed lines indicate crystallographic phase transitions as specified in the text. Insets to (b)-(d) display temperature evolution of the lattice orthorhombicity ($\eta = 2(a-b)/(a+b)$) for doped samples, with their maxima marked by vertical arrows. Vertical dashed lines indicate low temperature structural phase transitions. Sloping dashed straight lines are guides to the eye emphasizing anomalous trends seen in highlighted regions and discussed in the main text.}
\end{figure}

Typical Rietveld fits are shown in Fig.~\ref{fig:RietveldFitsQuality} for reference.  The resulting
temperature evolution of the in-plane lattice parameters is shown in Fig.~\ref{fig:lattice-fig}, where the vertical dashed lines indicate the temperatures of the structural phase transitions. These are in good agreement with published work~\cite{huck11}, as indicated in the phase diagram shown in Fig.~\ref{fig:AveragePhaseDiagram}.

\begin{figure}[tbf]
\includegraphics[width=80mm]{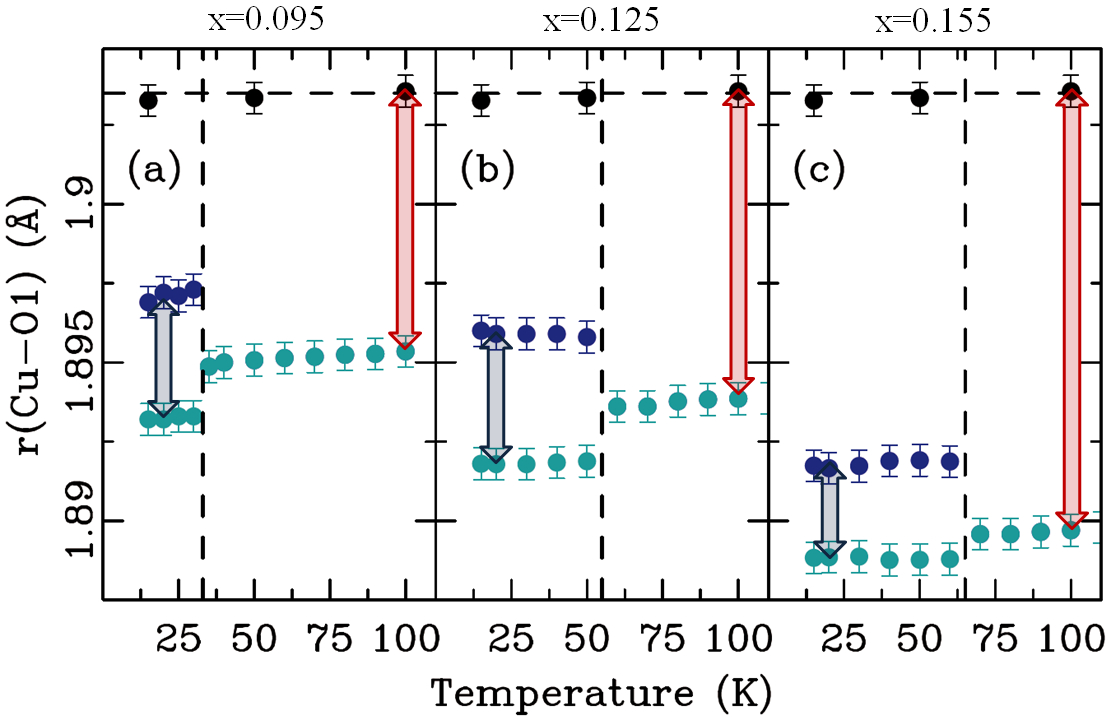}
\caption{\label{fig:C4C2-Rietveld-fig} (Color online) Temperature evolution of the average in-plane Cu-O bondlengths (solid blue circles) in \lbco\ obtained from Rietveld refinements: x=0.095 (a), x=0.125 (b), and x=0.155 (c). In all three panels: solid black symbols show in-plane Cu-O bond for x=0 sample as a reference. Horizontal dashed lines are guide to the eye, vertical dashed lines mark crystallographic phase transitions. Vertical double arrows highlight the magnitude of changes discussed in the main text.}
\end{figure}

Figure~\ref{fig:C4C2-Rietveld-fig} shows the evolution of the Rietveld refined average
in-plane Cu-O distances. Undoped \lco\ is in the LTO phase down to the lowest temperature and has a single
Cu-O planar-bond length (solid black circles in all panels of Fig.~\ref{fig:C4C2-Rietveld-fig}), highlighting the OE nature of the LTO tilts.
The biggest effect on doping is a significant shortening of the average Cu-O bond-length, highlighted by the red arrows in Fig.~\ref{fig:C4C2-Rietveld-fig}. This is due to the decrease in electronic charge in the Cu-O bonds, which are stabilized because charge is removed from anti-bonding states.  At low temperature, where the doped samples enter the LTT/LTLO phase \cite{huck11}, Rietveld analysis  reveals two distinct Cu-O bond-lengths consistent with the OI tilts.

It is noteworthy that
the difference in the in-plane Cu-O bondlengths of doped samples at base temperature is quite small---0.005(1)~\AA\ at most at 20~K, as indicated by the blue double arrows
in Fig.~\ref{fig:C4C2-Rietveld-fig}(a)-(c). This may be compared  to the distortions of $\sim$0.25~\AA\ observed in the ground state of \lmo\ due to the cooperative
Jahn-Teller effect~\cite{qiu;prl05,bozin;jpcs08}, which are nearly two orders of magnitude larger.  Nonetheless, because they are long-range ordered and can be observed crystallographically, such a small difference in the average bond-lengths can be reliably measured, giving us a direct indication of the broken symmetry between orthogonal Cu-O bonds.  Notably, despite the average tilt angles decreasing monotonically with increasing doping we see that the bond length mismatch (the difference between the $x$ and $y$ bond lengths) increases from $x=0.095$ to $x=0.125$ before decreasing again at $x=0.15$, suggesting a
stronger electronic stabilization of the orthogonal inequivalency around $1/8$ doping.

The first-order character of the low-temperature transition has a signature in the temperature dependence of the orthorhombic strain, shown in the insets to Fig.~\ref{fig:lattice-fig} (b)-(d).  On cooling through the LTO phase, the strain grows with decreasing temperature.  This growth slows as the low-temperature transition is approached so that at $\sim$30~K above the low-$T$ transition the average orthorhombicity even decreases, followed by a sharp drop at the transition.  The small decrease in strain on approaching the transition is consistent with the presence of a 2-phase coexistence region, as identified in past studies \cite{bill93,billi;pb94,huck11,wen12a}.

\begin{figure}[tbf]
\includegraphics[width=80mm]{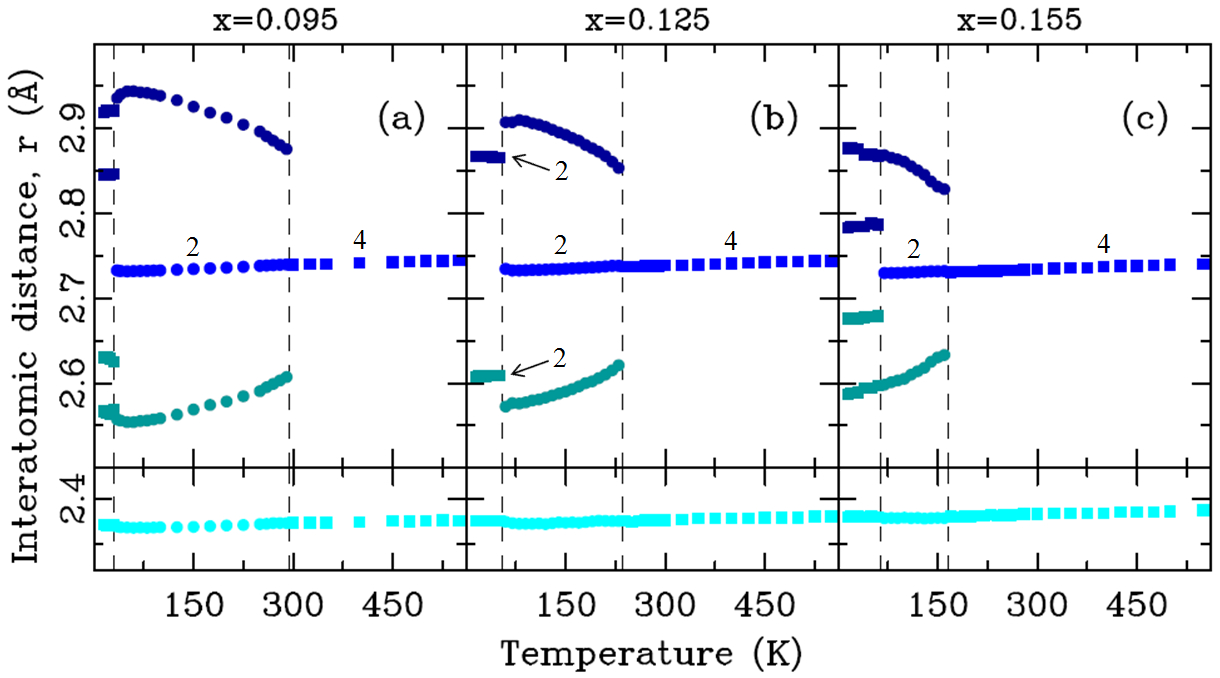}
\caption{\label{fig:LaOBonds} (Color online) Temperature evolution of the average La-O2 interatomic distances for (a) x=0.095, (b) x=0.125, and (c) x=0.155 composition. Bond multiplicity is indicated by a number where applicable. Short La-O2 bond that connects two LaO$_{2}$ planes is shown in separate panels. Vertical dashed lines indicate structural phase transitions.}
\end{figure}

If the LTO-HTT transition were purely displacive, then we would expect a continuous variation of structural parameters on passing through it.   Figure~\ref{fig:LaOBonds} demonstrates that this is not the case: the longer La-O2 bonds (O2 = apical oxygen) are split into three distances in the LTO phase, and this splitting abruptly drops to zero at the transition to the HTT phase.  Given the clear evidence that the transition is second order \cite{zhao07,huck11}, the apparent jump in the bond lengths must be an artifact of the Rietveld refinement.  The La-O2 bond length is affected most strongly by displacements of the apical oxygens, associated with octahedral tilts.  If there are disordered tilts present in the LTO phase, then the fitting process may result in unrealistic O2 displacements to compensate for the associated impact on Bragg peak intensities.  Note that in the fitting, we have assumed isotropic mean square atomic displacements, $U_{j}=\langle u_j^2\rangle$ where $u_j$ is the displacement of an atom at site $j$ and the average is over all equivalent sites in the sample. It follows that anisotropic fluctuations of the apical oxygens, such as occur in tilt fluctuations, might be modeled by finite displacements, even when the true average static displacement is essentially zero (just below the transition).

\begin{figure}[btf]
\includegraphics[width=80mm]{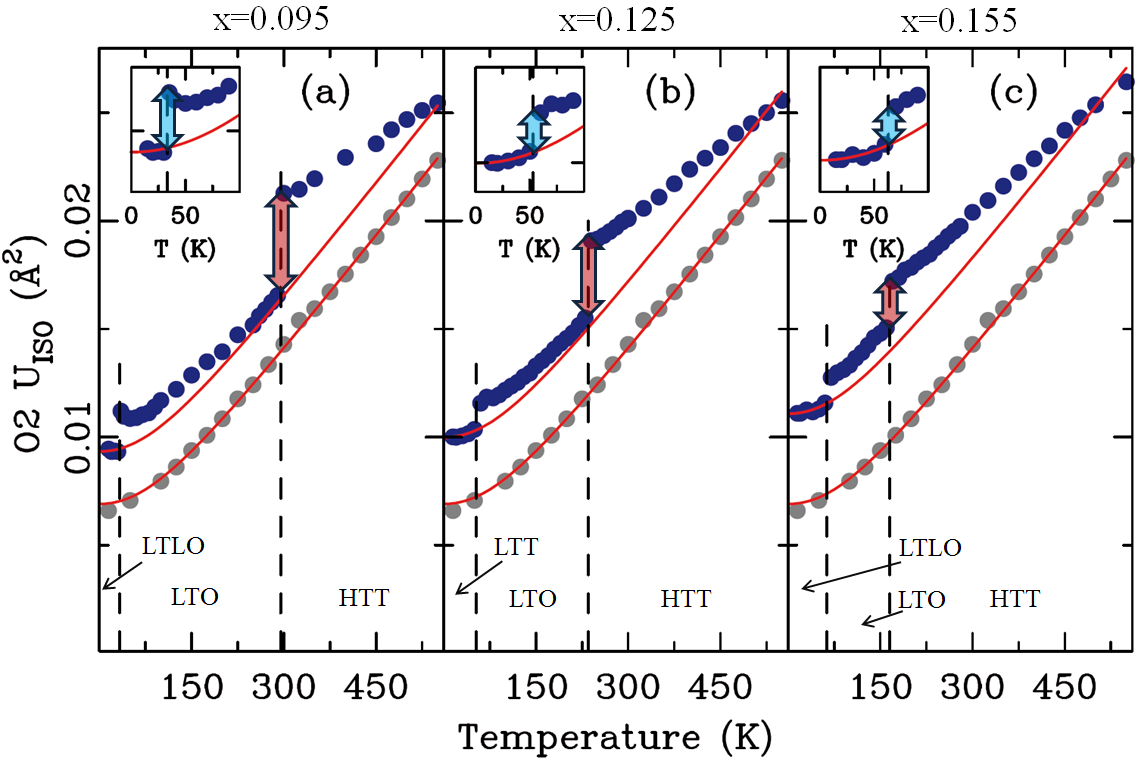}
\caption{\label{fig:uiso-fig} (Color online) Temperature evolution of atomic displacement parameter (ADP) of apical oxygen (solid blue symbols) in \lbco\ for: x=0.095 (a), x=0.125 (b), and x=0.155 (c). Vertical dashed lines indicate transitions between crystallographic phases as labeled and specified in the text. In all panels ADP data for x=0 sample are presented by solid gray symbols, with solid red line representing a fit of the Debye model, as discussed in the text. Debye model for ADPs of doped samples has the same Debye temperature as for x=0, but different offset such as to provide a good fit to ADP in the low temperature phases. Anomalous jumps in ADP discussed in the main text are indicated by double arrows. Insets focus on low temperature transitions.}
\end{figure}

If the disordered tilts change continuously across the transition, then we should expect to see an anomalous jump in $U_{iso}$ for O2 on entering the HTT phase.  Figure~\ref{fig:uiso-fig} shows that this, indeed, is the case.  In fact, there is also a jump at the low-temperature transition, indicating that there are anisotropic fluctuations present in the LTO phase that cannot be compensated by adjusting the symmetry-allowed atomic displacements.  The lower red curve corresponds to a Debye-model fit \cite{debye;adp12} to the O2 $U_{iso}$ of La$_2$CuO$_4$ (gray circles).  This involves two parameters, the Debye temperature $\theta_D$, and an offset factor $U_0$ \cite{knox;prb13,bozin;sr14}; the behavior for La$_2$CuO$_4$ is well described by the parameter values $\theta_{D}$ of 500(2)~K and $U_{0}$ of 0.0023(2)~\AA$^{2}$.  In the other Figure panels, in each case the upper red curve is obtained by adding a doping-dependent offset to match the experimental values at the lowest temperature, but keeping the Debye temperature the same as that obtained for the undoped end-member.  The offset grows with Ba concentration, which is consistent with the disorder in O2 positions induced by Ba substitution for La \cite{hask00}.  It appears to capture the underlying trend in each case, with the excess disorder associated with the jumps gradually becoming indistinguishable from the thermal effects with increasing temperature.  This is in agreement with earlier observations in \lbco\ and \lsco\ of persisting local tilt fluctuations across these macroscopic
phase transitions~\cite{billi;prl94,bozin;ssp98,bozin;pb98,bozin;prb99}.
Similar anomalies have been seen in other systems, where they are typically ascribed either to the inadequacy of the structural model used, or sometimes more specifically to the presence of
nanoscale structural features, such as broken symmetry states, that do not propagate over long length-scales~\cite{bozin;s10,knox;prb13,knox;prb14,bozin;sr14}.

%
\subsection{Local structure}
\label{localstructure}
\begin{figure}[tbf]
\includegraphics[width=80mm]{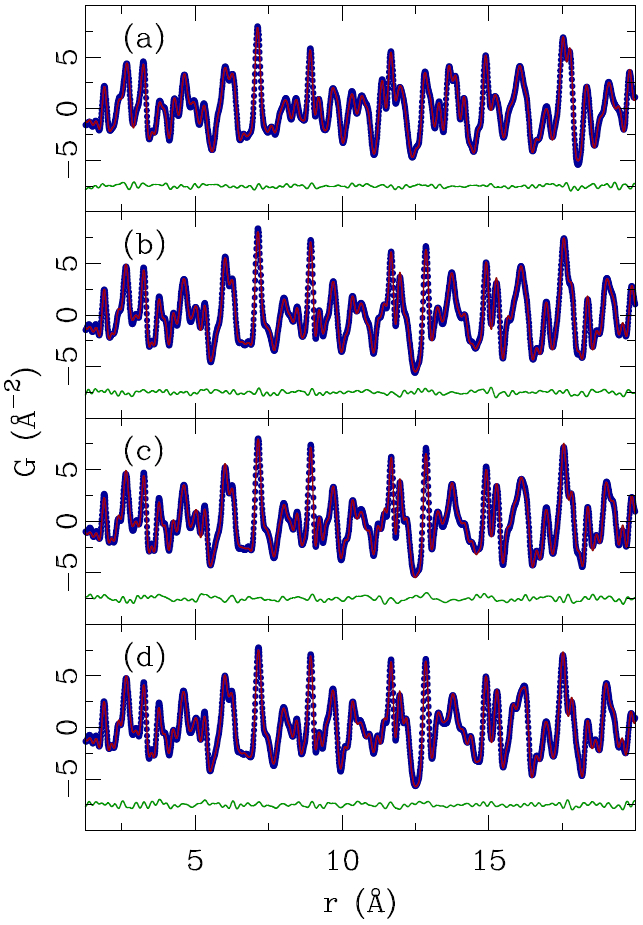}
\caption{\label{fig:PDFFitsQuality} (Color online) PDF fits of the average structure models to LBCO data at 15~K. Closed blue symbols represent the data, solid red lines are the models, and solid green lines are the differences (offset for clarity). (a) x=0 using \bmab\ model, (b) x=0.095 using \pccn\ model, (c) x=0.125 using \p42ncm\ model, and (d) x=0.155 using \pccn\ model.}
\end{figure}

%
Using the same neutron scattering data, we have characterized the local structure of \lbco\   by the PDF approach.  Figure~\ref{fig:PDFFitsQuality} shows PDF fits over intermediate $r$-ranges at 15~K using the average structure models, establishing the overall data quality and displaying that good fits can be obtained on this length scale.  Here, we have taken account of the correlated motion of short interatomic bonds~\cite{jeong;jpc99,proff;jac99} by defining the mean-square relative displacement $\sigma_{ij}^2$ of atoms at positions $i$ and $j$, separated by distance $r_{ij}$, as
\begin{equation}
  \sigma_{ij}^2 = (\langle u_i^2\rangle + \langle u_j^2\rangle)[1-(r_0/r_{ij})],
\end{equation}
where the parameter $r_0$ is fixed at 1.6~\AA\  for all PDF calculations in this paper.

\begin{figure}[tbf]
\includegraphics[width=80mm]{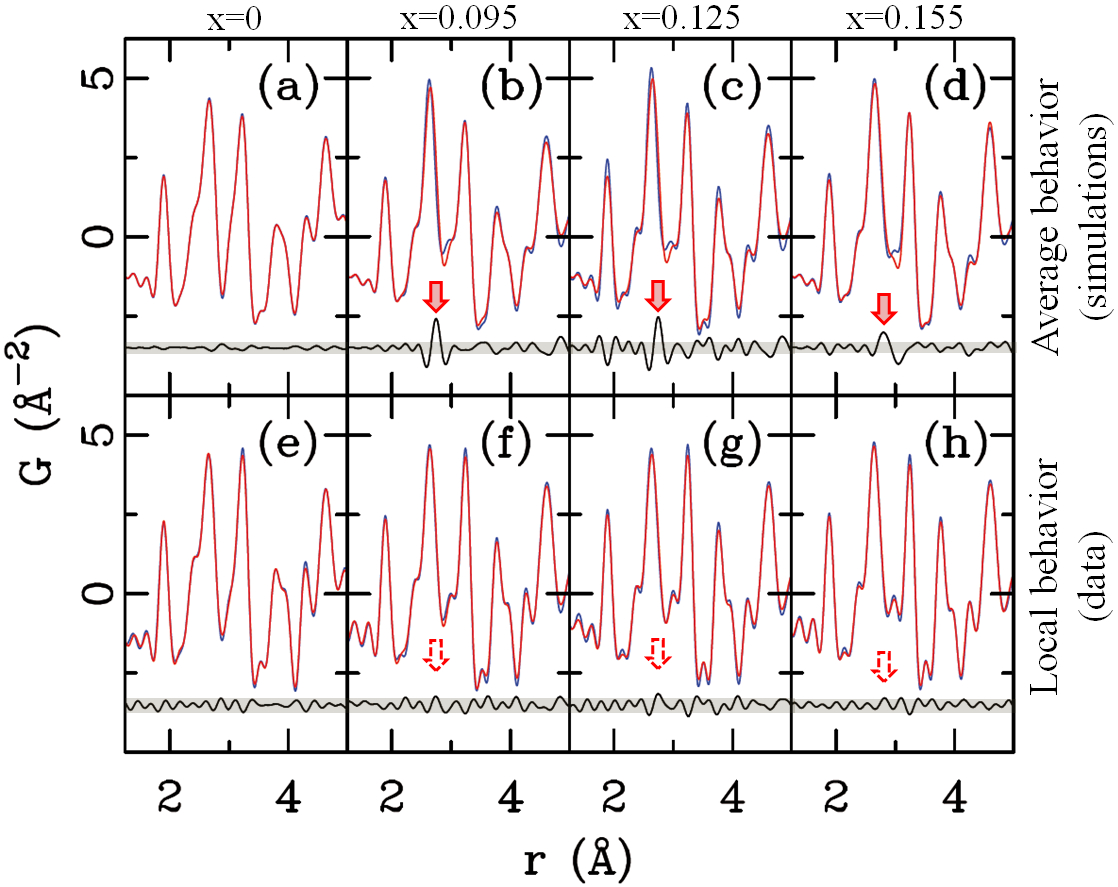}
\caption{\label{fig:Average-vs-Local-PDF-fig} (Color online) PDF comparison of the average and local structure behavior in \lbco. Top row: simulated PDFs calculated using parameters from fully converged Rietveld refinements reflecting average structure behavior. Low-T profiles are shown in blue, higher-T profiles are shown in red. (a) 15~K vs 50~K LTO structure PDFs for x=0; (b) 15~K LTLO vs 60~K LTO PDFs for x=0.095; (c) 15~K LTT vs 80~K LTO PDFs for x=0.125; (d) 15~K LTLO vs 70~K LTO PDFs for x=0.155. Bottom row: comparison of the raw experimental PDF data for the same respective temperatures as considered in (a)-(d); lower temperature data shown in blue, higher temperature data shown in red. Difference curves (high-T minus low-T PDF) are offset for clarity. Shaded areas represent span of the difference curves observed for x=0 composition. Changes observed in the average structure (marked by filled arrows) are not observed in the local structure (empty arrows), as experimental PDFs do not change across the OE/OI phase transitions in doped \lbco\ samples.}
\end{figure}

To explore the origin of enhanced mean-square displacements in the Rietiveld refinements shown in Fig.~\ref{fig:uiso-fig}, we focus on the very local structure.
We first evaluate the expected effects on the PDF of the symmetry change across
the low-temperature transition in the case where the local and average structures agreed. Fig.~\ref{fig:Average-vs-Local-PDF-fig}(b)-(d) shows
a comparison of PDFs simulated using parameters from the Rietveld refinements at base temperature (blue profile, OI model) and at the temperature of maximum orthorhombic strain
(red profile, OE model), bracketing the low temperature transition in the doped samples.  The changes expected in the PDF across the transition are clearly observable in the respective difference curves, with the strongest features just below 3~\AA\ (marked by arrows), corresponding to La--O2 bonds.

The actual measured PDFs at the same temperatures are shown in Fig.~\ref{fig:Average-vs-Local-PDF-fig}(f)-(h).  The difference curves show nothing above the noise level, indicating the absence of change in the local structure across the transition, in contrast to the predictions from the average structure.
In fact, the temperature difference is similar to that observed in pure \lco, Fig.~\ref{fig:Average-vs-Local-PDF-fig}(e), where there is no change in the average structure.   These results are consistent with earlier work \cite{billi;prl94}.

\begin{figure}[tbf]
\includegraphics[width=80mm]{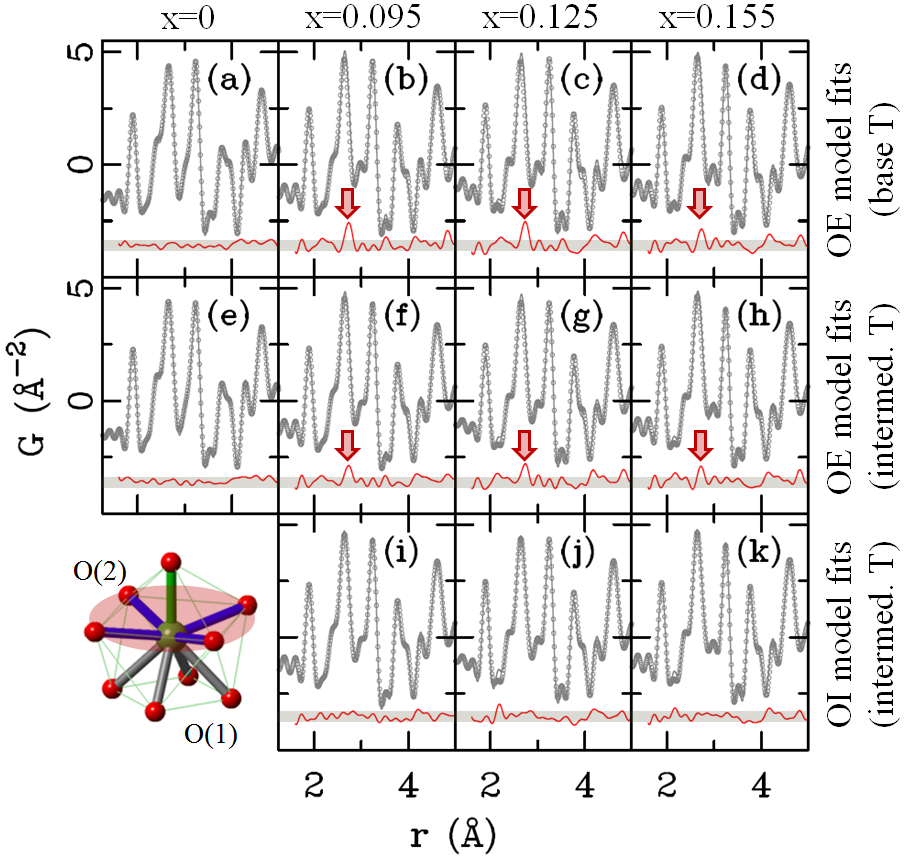}
\caption{\label{fig:Local-PDF-data-vs-models-fig} (Color online) OE versus OI PDF models for the local structure (solid gray lines). Data are shown as open gray symbols, difference curves (solid red lines) are offset for clarity. (a)-(d) 15~K data vs LTO model (OE symmetry) – explains x=0 data well (as it should), fails for x=0.095, 0.125, and 0.155 (as it should). The largest discrepancies are marked by arrows. (e) LTO for x=0 sample at 50~K, (f)-(h) data at T of maximum orthorhombicity (60~K, 80~K, and 70~K for x=0.095, 0.125, and 0.155 respectively) vs LTO model (fails at same places as at 15~K); (i)-(k) same as (f)-(h) but with LTLO/LTT models (OI symmetry) –- underlying data correspond to OI symmetry. Inset: La/Ba-O(2) distances (shaded area) contribute principally to the misfits marked by arrows in (b)-(d) and (f)-(h).}
\end{figure}

Further confirmation of this comes from explicit short range PDF modeling that was carried out using both OE-type and OI-type models fit to the data at
base temperature and at the maximum orthorhombicity temperature in the doped samples. These fits are shown in Fig.~\ref{fig:Local-PDF-data-vs-models-fig}.
While the OE-type model readily explains the \lco\ data at both base and intermediate temperature, Fig.~\ref{fig:Local-PDF-data-vs-models-fig}(a) and (e), it gives an unsatisfactory fit to the region of the La--O2 bonds in the doped samples. Such a discrepancy at base temperature (Fig.~\ref{fig:Local-PDF-data-vs-models-fig}(b)-(d)) is expected, since the
underlying atomic structure there is OI.  At the temperature of maximum orthorhombic strain, the discrepancy with the OE fits remains, as shown in
Fig.~\ref{fig:Local-PDF-data-vs-models-fig}(f)-(h), whereas OI fits do much better, as indicated in Fig.~\ref{fig:Local-PDF-data-vs-models-fig}(i)-(k).

\begin{figure}[tbf]
\includegraphics[width=80mm]{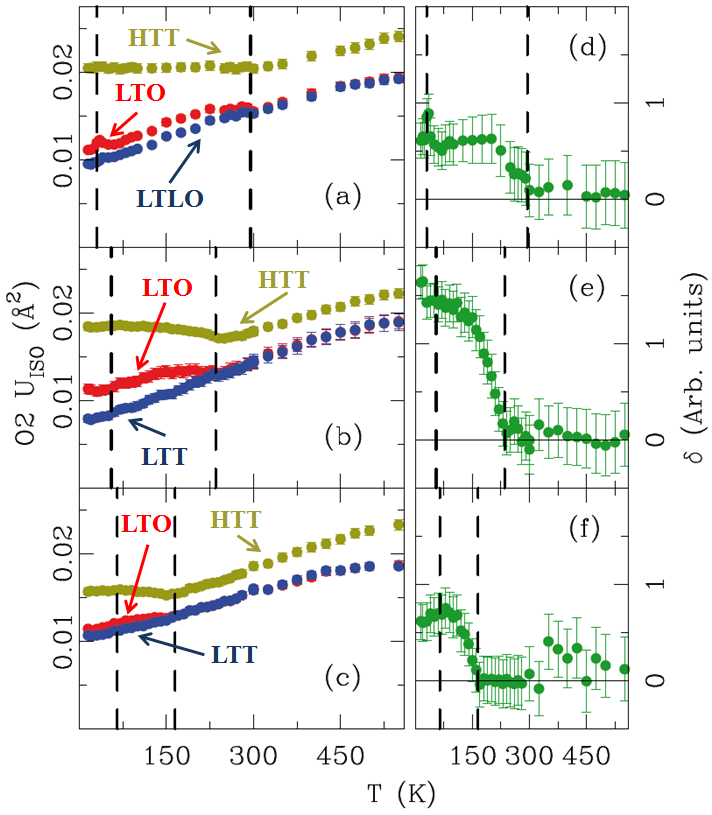}
\caption{\label{fig:PIso-ADP-O2} (Color online) Semi-quantitative exploration of the OI-ness of LBCO via assessment of T-dependence of apical oxygen isotropic atomic displacement parameter (ADP) as obtained by fitting different structural models (as indicated by arrows in (a)-(c)) to the PDF data over 4~nm $r$-range. (a) x=0.095, (b) x=0.125, and (c) x=0.155. Difference between the observed ADPs using LTO and LTLO/LTT models from panels (a)-(c) are displayed in (d)-(f) for x=0.095,  x=0.125, and x=0.155 respectively, normalized by the square of the average transverse displacement of apical oxygen. See text for details.}
\end{figure}

We saw previously that $U_{iso}({\rm O}(2))$ provides a distinctive measure of the tilt disorder that cannot be simulated by the symmetry-allowed structural parameters in the Rietveld refinements.  We now consider the behavior of $U_{iso}({\rm O}(2))$ obtained from fits to the PDF data (for the range $15 < r < 40$ \AA), as shown in Fig.~\ref{fig:PIso-ADP-O2}(a)--(c).  At each temperature, separate fits have been performed with the LTT (LTLO), LTO, and HTT models.  At all temperatures, we find that the $R$-factor, measuring the quality of fit, is always smallest for LTT (LTLO), followed by LTO, and then HTT.   As one can see, $U_{iso}({\rm O}(2))$ from the LTT (LTLO) fit shows a monotonic increase with temperature, with no anomalies at the transition temperatures.  The LTO fit is consistent with LTT at high temperature, but is larger in the LTO and LTT phases.  The results for HTT are considerably larger at all temperatures.

As a measure of the distinctly OI tilts, we define the parameter $\delta$ as
\begin{equation}
  \delta=\Delta U_{iso}/\langle x_{\rm O(2)} \rangle ^{2},
  \label{eq:delta}
\end{equation}
where $\Delta U_{iso}$ is the difference in parameter values obtained from the LTO and LTT (LTLO) fits, normalized to the square of the average transverse displacement of the O(2) site in the LTT phase at low temperature,  $\langle x_{\rm O(2)} \rangle^2$.  The temperature dependence of $\delta$ is plotted in Fig.~\ref{fig:PIso-ADP-O2}(d)--(f) for the doped samples.  For $x=0.125$, we find evidence for substantial OI tilts throughout the LTO phase, with a reduced magnitude for $x=0.095$ and 0.155.  On entering the HTT phase, we have already seen evidence that tilt disorder is present; however, the fact that $\delta\approx 0$ suggests that there is little preference between OI and OE tilts at high temperature.  It is consistent with the idea that, in the HTT phase, the Cu-O(2) bond precesses rather smoothly around the $z$ axis as discussed before~\cite{axe89,pick91,isaac;prl94}.

\begin{figure}[tbf]
\includegraphics[width=80mm]{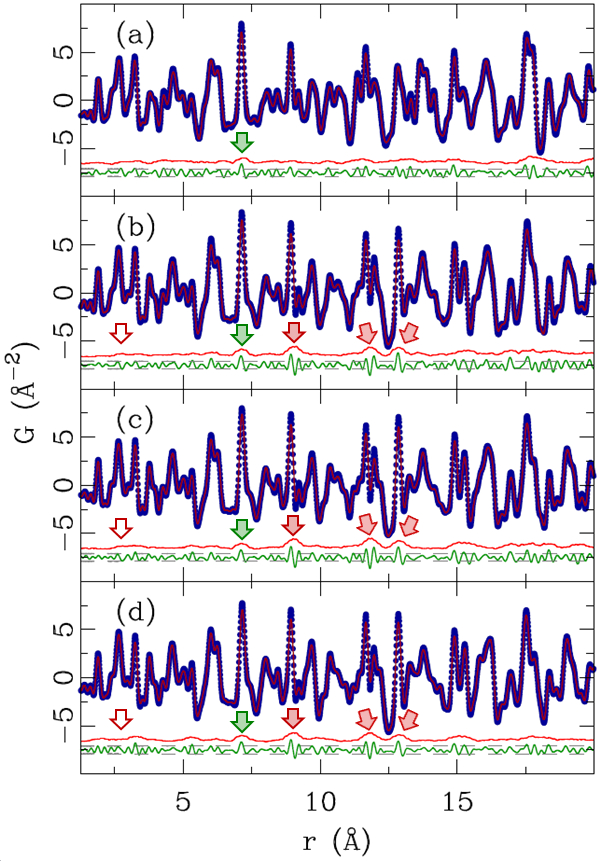}
\caption{\label{fig:PDFdataComparison-15vs100} (Color online) Comparison of LBCO PDF data at 15~K (closed blue symbols) and at 100~K (solid red line) with difference curve (solid green line) offset for clarity. Horizontal gray dashed lines mark experimental uncertainty on 2$\sigma$~level. Light-red solid line represents a 0.5~\AA\ running average of the absolute value of the difference curve, multiplied by 2 and offset for clarity. (a) x=0, (b) x=0.095, (c) x=0.125, and (d) x=0.155. Difference in data for x=0 sample displays the change expected from canonical thermal evolution effects without symmetry changes. Differences in the data for all three doped samples display similar level of change as x=0 sample up to $\sim$ 9~\AA\, with first substantial changes seen on longer length-scale. Low $r$ assessment: green arrows mark changes seen in the data for all the samples, while the red arrows mark significant changes seen only in the doped samples, presumably associated with the change in average symmetry. Empty arrows around 3~\AA\ mark places where changes are expected from the average structure, but not observed locally, as shown in Fig.~\ref{fig:Average-vs-Local-PDF-fig} and discussed in the text. Local structure across the global OI to OE phase transition is preserved on sub-nanometer length-scale.}
\end{figure}

We now address the length-scale of the local OI tilts.
This can be obtained qualitatively from a direct comparison of
the experimental PDF data at 15~K and 100~K (well above the phase coexistence region) for all samples studied. This is shown in Fig.~\ref{fig:PDFdataComparison-15vs100}(a)-(d).
Panel (a) shows the result for \lco;  only small differences are seen in the PDFs beyond the expected uncertainties (e.g., indicated by the green arrow), and these provide a measure of the signal we should expect due to thermal effects within the same phase.
From Fig.~\ref{fig:PDFdataComparison-15vs100}(b)--(d), it is clear that in the doped samples there are additional signals in the difference curve, starting near a distance of 9~\AA , that are reproducible from sample to sample and especially evident in the running average of the difference curve (red curve).  This suggests that the OI tilt correlations within the LTO phase have a correlation length that is $<9$~\AA.

The PDF analysis presented here is based on total scattering data that do not discriminate between the elastic and inelastic scattering channels, and hence the PDF does not
distinguish whether the underlying short-range features are static or dynamic.  This is in contrast to the Rietveld analysis, which is sensitive only to the time-averaged information in the Bragg peaks; the inelastic information is largely in the tails of the Bragg peaks.  To test the static or dynamic character, we turn next to inelastic neutron scattering on a single crystal, focusing on \lbco\ with $x=0.125$.

%
\subsection{Octahedral tilt dynamics}
\label{tiltdynamics}
%

\begin{figure}[tbf]
\includegraphics[width=80mm]{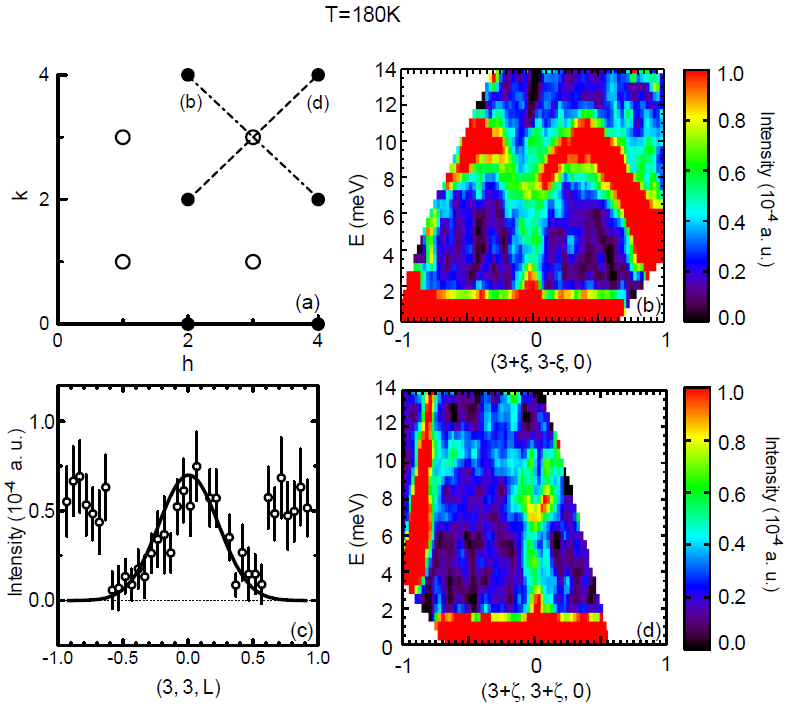}
\caption{\label{fig:INS-01} (Color online) (a) Diagram of the $(H,K,0)$ plane of reciprocal space indicating fundamental Bragg peaks (filled circles) and LTT superlattice peaks (open circles), with dot-dashed (dashed) line indicating the orientation of the data slice in (b) [(d)].  (b) Map of scattering intensity for $E$ vs.\ ${\bf Q}=(3+\xi,3-\xi,0)$.  (c) Intensity (integrated over $ 2 \le E \le 4~meV$) vs.\ ${\bf Q}=(3,3,L)$.  (d) Intensity map for $E$ vs.\ ${\bf Q}=(3+\zeta,3+\zeta,0)$.  All measurements are at $T=180$~K, in the LTO phase.}
\end{figure}

\begin{figure}[t]
\includegraphics[width=80mm]{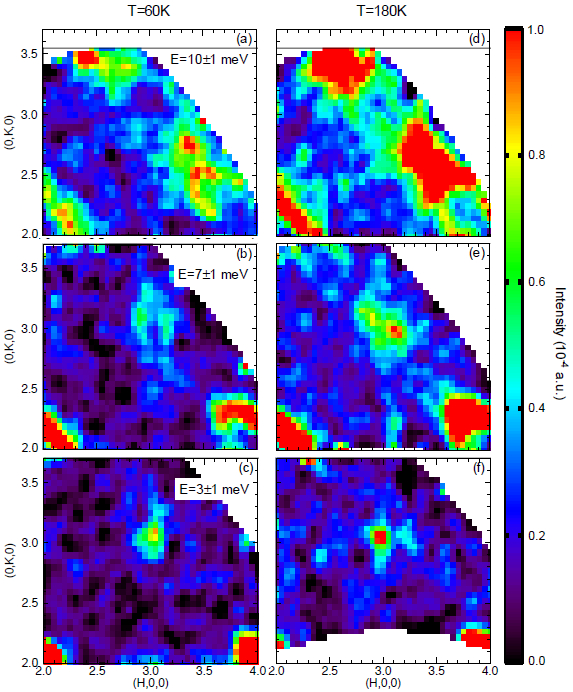}
\caption{\label{fig:INS-02} (Color online) Constant-energy slices, with signal integrated over $\pm1$~meV, through the soft-phonon scattering around (330). (a,d) $E=10$~meV; (b,e) $E=7$~meV; (c,f) $E=3$~meV.  Data obtained at $T=60$~K for (a--c) and 180~K for (d--f).}
\end{figure}

The inelastic scattering about the (330) reciprocal point of our \lbco\ with $x=0.125$ is shown in Fig.~\ref{fig:INS-01}. The cuts in reciprocal space
that were taken are shown schematically in Fig.~\ref{fig:INS-01}(a).
Figures~\ref{fig:INS-01}(b) and (d) show the dispersion of excitations along the transverse and longitudinal directions, respectively, within the LTO phase at $T=180$~K.  In both cases, one can see a soft phonon with intensity that can be resolved between 2 and 10~meV.  In the transverse direction, these excitations connect to the transverse acoustic modes dispersing from the neighboring (240) and (420) fundamental Bragg points.  Another perspective is given by the constant energy slices shown in Fig.~\ref{fig:INS-02} for several different energies, where we compare with results at 60~K, slightly above the low-temperature transition.  For dispersion in the longitudinal direction, the intensity becomes quite weak as one moves away from the (330) point.  Figure~\ref{fig:INS-01}(c) shows that the excitations, integrated between 2 and 4 meV, have a finite width along $Q_z$, demonstrating that the LTT tilt fluctuations have 3D character.

\begin{figure}[tbf]
\includegraphics[width=80mm]{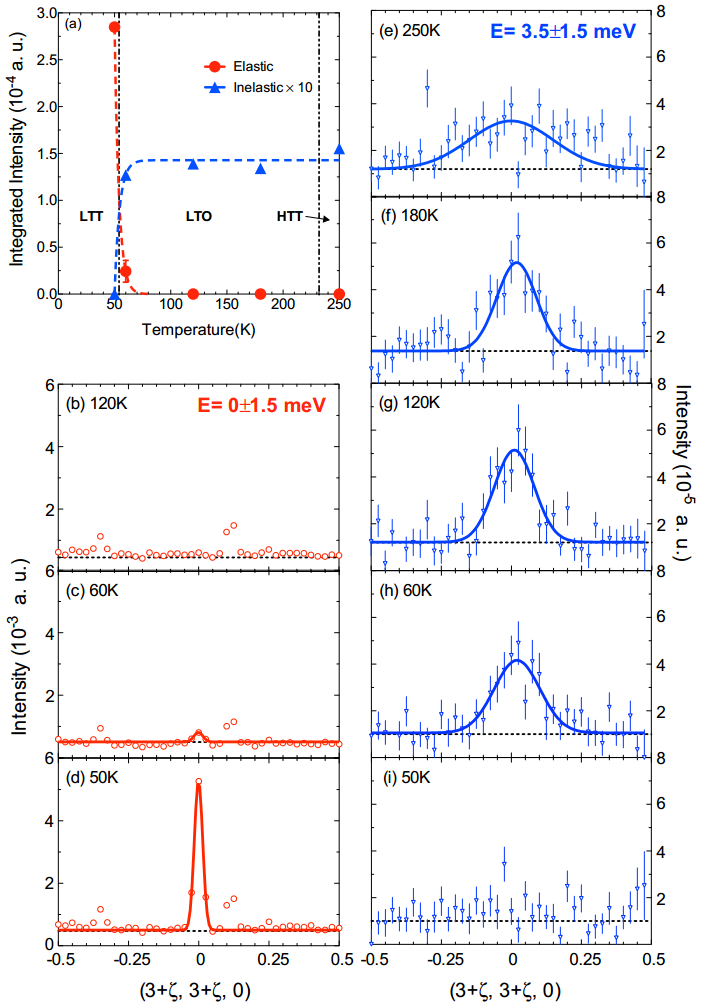}
\caption{\label{fig:INS-results-temporary-LBCO0125-fig} (Color online) Single crystal results at the (330) LTT superlattice position for $x=0.125$. (a) Summary of the temperature dependence of the elastic (red circles) and inelastic (blue triangles) integrated intensities obtained from the following panels; vertical dashed lines denote phase boundaries, while dashed lines through data points are guides to the eye. (b)-(d) Elastic channel (integrated over $\pm1.5$ meV) measured along the longitudinal direction at 120, 60, and 50~K, respectively. Solid lines are Gaussian-peak fits, used to determine the integrated intensity; weak, $T$-independent peaks are diffraction from the aluminum sample holder. (e)-(i) Inelastic signal from the soft-phonon fluctuations (2-5 meV integration) measured at 250, 180, 120, 60, and 50 K, respectively.  Lines are Gaussian-peak fits.}
\end{figure}

The temperature dependence of the scattering near (330) is presented in Fig.~\ref{fig:INS-results-temporary-LBCO0125-fig}.   There is a clear superlattice reflection at (330) in the LTT phase at 50~K.  Warming to 60 K, just across the transition to the LTO phase, very weak elastic scattering is still detectable; however, this is completely gone at 120~K.  In contrast, soft phonon fluctuations (integrated over 2 to 5 meV) centered at (330) are clearly seen in the LTO phase and even in the HTT, at 250 K.  At 50 K, the intensity from the soft fluctuations has all condensed into the elastic superlattice peak.  (There must be acoustic phonons dispersing out of the superlattice peak, but these are too weak for us to detect.)  The temperature dependences of both the elastic and inelastic signals are summarized in Fig.~\ref{fig:INS-results-temporary-LBCO0125-fig}(a).

The correlation length for the LTT-like tilts within the LTO phase can be estimated from the $Q$-width of the soft-phonon scattering.  Taking the inverse of the half-width-at-half-maximum for the fitted peaks in Fig.~\ref{fig:INS-results-temporary-LBCO0125-fig}(e-h), we find an effective correlation length of $\sim5$~\AA\ in the LTO phase, decreasing to about half of that in the HTT phase.  This is consistent with estimate of $<9$~\AA\ obtained from the PDF analysis.

\begin{figure}
\includegraphics[width=80mm]{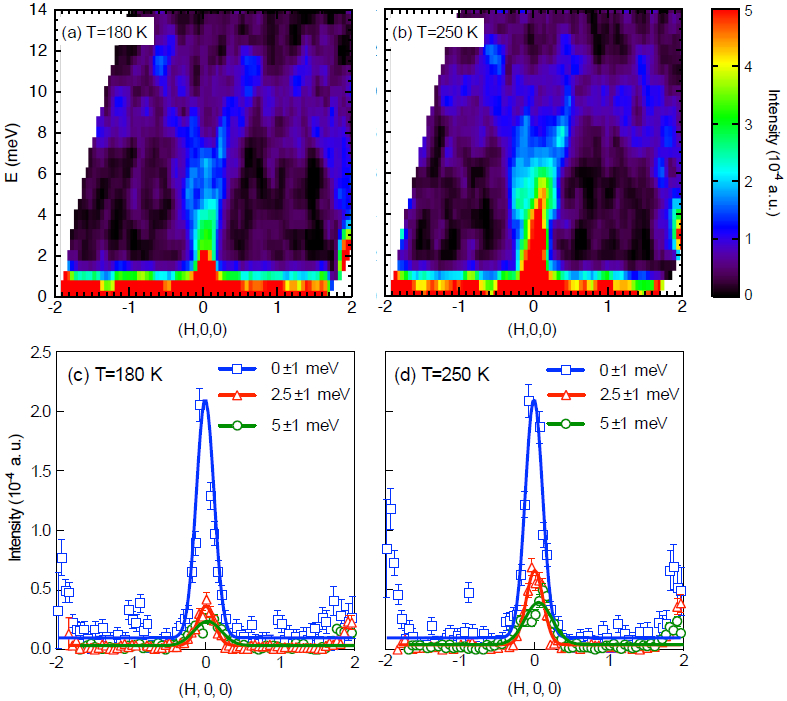}
\caption{\label{fig:INS-05} (Color online) (a,b) Intensity map as a function of energy vs.\ ${\bf Q}=(H,3,2)$, showing the transverse dispersion of tilt modes about the (032) superlattice peak position of the LTO phase, obtained at $T=180$ and 250~K, respectively.  (c,d) Intensity, integrated over a window of $\pm 1$~meV along the transverse direction, for $E=0$ (blue squares), 2.5 (red triangle), and 5 meV (green circles), at $T=180$ and 250~K, respectively.}
\end{figure}

For comparison, Fig.~\ref{fig:INS-05} (a) and (b) show the dispersion of tilt fluctuations in the transverse direction about the (032) position (an LTO superlattice peak) in the LTO and HTT phases, respectively.  The intensity is much stronger than at (330) because of a much larger structure factor.   There is substantial intensity from soft tilt fluctuations, and even quasielastic scattering, at 250~K in the HTT phase, as seen previously \cite{kimu05}.  In the LTO phase at 180 K, much of the low-energy weight is due to the residual soft mode that condenses at the LTT transition.   We note that scattering at (032) is allowed in both the LTO and LTT phases; it follows that one cannot uniquely distinguish between OE and OI soft tilt fluctuations at this $\vec{Q}$-point, in contrast to fluctuations at (330).

%
\section{Discussion}
\label{discussion}
%

\begin{figure}[tbf]
\includegraphics[width=50mm]{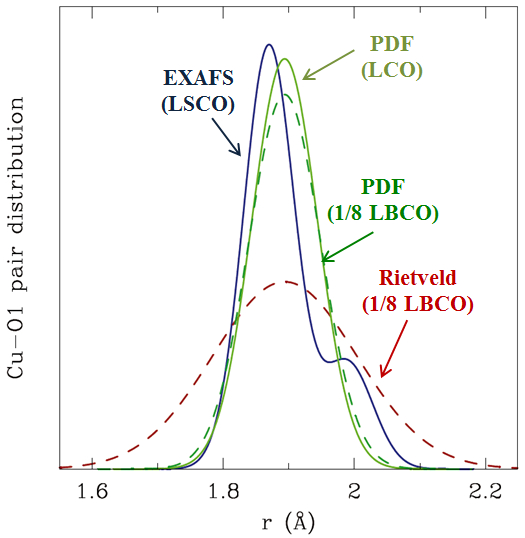}
\caption{\label{fig:PairDistribution-EXAFS} (Color online) Comparison of measures of the in-plane Cu-O bond-length distribution. Green lines show the profile for \lbco\ with $x=0$ (solid) and $x=0.125$ (dashed).  The effective profile for the latter sample from the Rietveld refinement (red dashed line) is not sensitive to correlated motion.  The blue solid line indicates the profile for \lsco\ with $x=0.15$ and $T=15$~K from Bianconi {\it et al.} \cite{bian96}.
}
\end{figure}

The distribution of Cu-O nearest-neighbor bond lengths has been the subject of some controversy over the years, and so deserves some discussion.  Our Rietveld analysis finds a maximum bond length splitting of just 0.005~\AA\ in the LTT phase of \lbco\ with $x=0.125$, consistent with earlier diffraction work on \lbco\  \cite{cox89}.  To appreciate how small this splitting is, we compare various measures of the bond-length distribution in Fig.~\ref{fig:PairDistribution-EXAFS}.  The peak obtained from the PDF analysis yields a mean-squared relative displacement, $\sigma^2$, of 0.0022(1)~\AA$^2$ at 15~K, corresponding to a bond-length spread of 0.05~\AA; hence, the disorder in the bond length, largely due to zero-point fluctuations, is an order of magnitude greater than the bond-length splitting.  The width of the PDF peak is significantly smaller than that obtained from the Rietveld analysis, as the former is sensitive to the correlated motion of nearest neighbors, whereas diffraction intensities only have information on the independent fluctuations of the distinct atomic sites.

The Cu-O pair distribution can also be probed in x-ray absorption fine structure (XAFS) studies.  Bianconi and coworkers \cite{bian96,sain97} have reported a splitting of the Cu-O bond distribution by 0.08~\AA\ below 100~K in La$_{1.85}$Sr$_{0.15}$CuO$_4$.  We have reproduced their low-temperature distribution in Fig.~\ref{fig:PairDistribution-EXAFS}.  In a study of La$_{1.875}$Ba$_{0.125}$CuO$_4$ \cite{lanz96}, they reported a corresponding anomalous increase of $\sigma^2$ by 0.001~\AA$^2$ on cooling below 60 K.   Such observations are incompatible with our results.  As can be seen in Fig.~\ref{fig:PairDistribution-EXAFS}, the PDF analysis has sufficient resolution to detect the splitting claimed in the \lsco\ study \cite{bian96,sain97}.  Regarding the temperature dependence of $\sigma^2$, we find on warming to 100 K that there is small increase to 0.0025(1)~\AA$^2$, not a decrease.  Our finding of an absence of anomalous bond length disorder at low temperature is consistent with that of other XAFS studies \cite{hask00,tran87b}.  It is also consistent with estimates of the bond-length modulation associated with stripe order based on superlattice intensities \cite{tran96b}.  We conclude that the bond-length modulation that pins stripe order is remarkably subtle.

\begin{figure}[tbf]
\includegraphics[width=80mm]{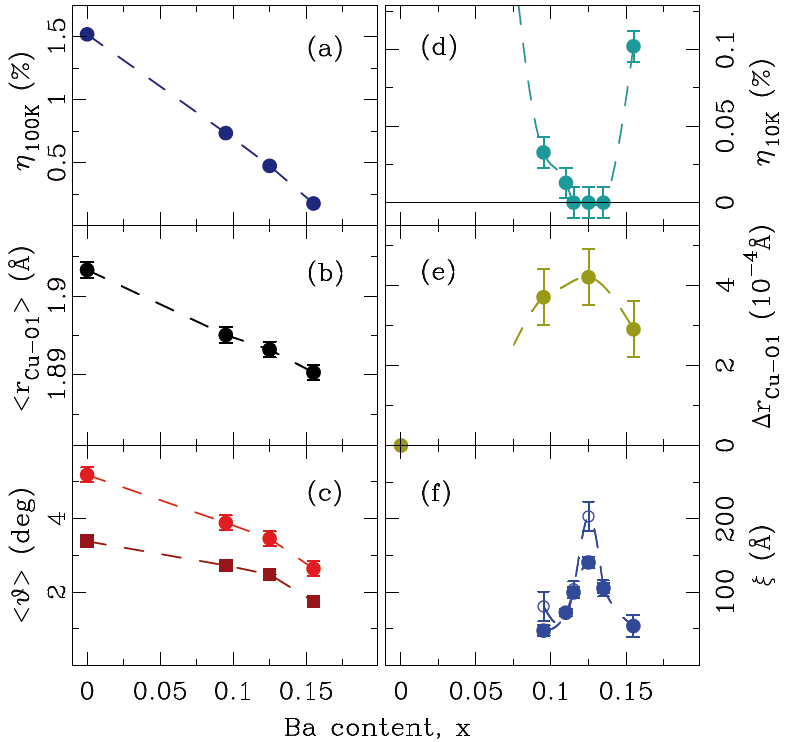}
\caption{\label{fig:MonotonicNonmonotonic} (Color online) Doping dependence of various system parameters of \lbco. (a) Orthorhombicity at 100~K (dark blue solid symbols) from Rietveld. (b) Average planar Cu-O bond distance (solid black symbols) at 100~K temperature. (c) Average \cuosix\ tilt angle extracted from apical oxygen (light red solid circles) and planar oxygen (dark red solid squares) positions at maximum orthorhombicity. (d) Orthorhombicity at 10~K after {H\"ucker} et al.~\cite{huck11} (light blue solid symbols). (e) Planar Cu-O bondlength anisotropy at 15~K (olive solid symbols). (f) In-plane correlation lengths $\xi$ of charge ordering parallel (solid blue circles) and perpendicular (open blue circles) to the stripe direction at base temperature~\cite{huck11}.
}
\end{figure}

Next, we turn to the doping dependence of the Cu-O bond anisotropy, where we have two competing trends.  One of these involves the decrease in the average octahedral tilt with doping.  The shortening of the in-plane Cu-O bond length reduces the mismatch with bond lengths in the La$_2$O$_2$ layer, resulting in one contribution to the reduction in average tilt.  Another comes from the quenched disorder associated with substituting Ba$^{2+}$ for La$^{3+}$.  The Ba acts effectively as a negative defect, repelling the neighboring apical oxygens and disrupting the octahedral tilt pattern.  These effects lead to the decrease in the average orthorhombic strain with $x$, as summarized in Fig.~\ref{fig:MonotonicNonmonotonic}.  The competing trend involves the onset temperature for the LTT (LTLO) transition, resulting in ordering of OI tilts.  The empirical trend is that this should grow with Ba concentration (as it depends on the average ionic radius of the $2+$ ions relative to that of the $3+$ ions \cite{suzu94}).  These competing trends lead to the bond-length anisotropy reaching a maximum that appears to occur coincidentally near $x=0.125$, where stripe order is strongest \cite{huck11}.

\begin{figure}[tbf]
\includegraphics[width=80mm]{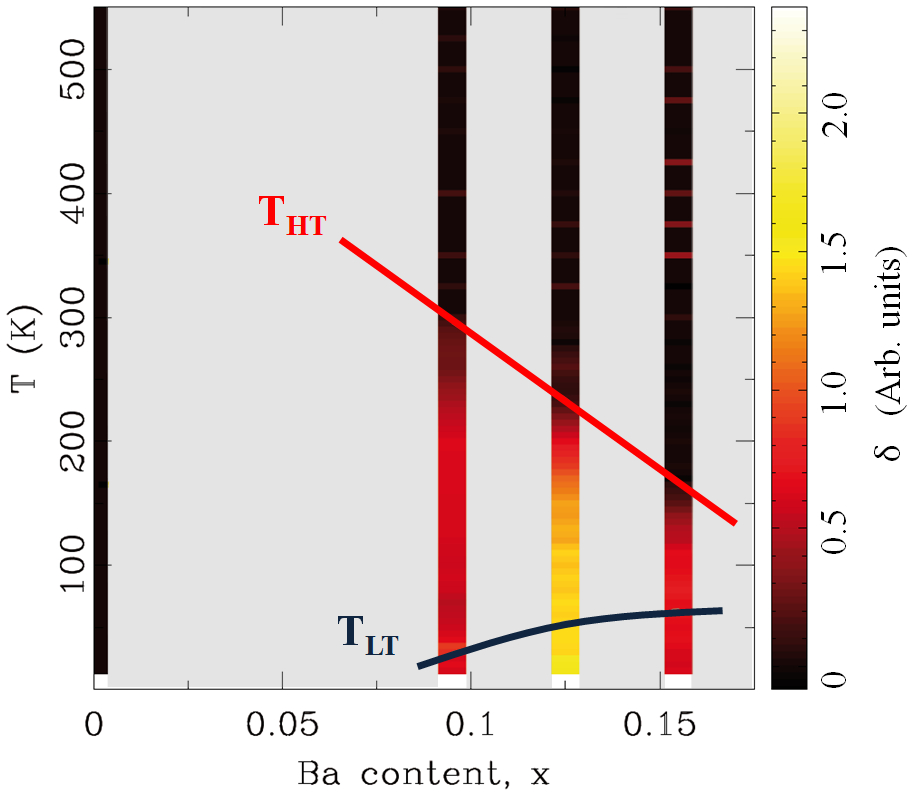}
\caption{\label{fig:FalseColorDiagram} (Color online) (x, T) evolution of $\delta$ shown in Fig.~\ref{fig:PIso-ADP-O2}(d)-(f). Solid lines mark structural transition temperature: T$_{HT}$ (red) and T$_{LT}$ (blue).}
\end{figure}

We get a somewhat different perspective from the temperature and doping dependence of parameter $\delta$, defined in Eq.~(\ref{eq:delta}), which measures the LTT-like component of the octahedral tilts; a false color representation is shown in Fig.~\ref{fig:FalseColorDiagram}.  Here we see that the OI tilts appear in a significant way at the onset of LTO order.  Furthermore, the relative magnitude is largest at $x=0.125$ even before LTT order is achieved.   This behavior is suggestive that the structural anisotropy might be influenced by electron-phonon coupling and particular stability of the stripe-ordered phase at $x=0.125$.  Regarding  stability, it is interesting to note the results of recent optical pump probe studies.   Pumping the $x=0.125$ phase in La$_{1.8-x}$Eu$_{0.2}$Sr$_x$CuO$_4$ with 80-meV photons with polarization {\it parallel} to the planes induces interlayer superconducting coherence for $T\lesssim15$~K \cite{faus11}; this also causes stripe melting in \lbco\ \cite{fors14}.  In contrast, pumping \lbco\ with polarization {\it perpendicular} to the planes enhances interlayer superconducting coherence for $x=0.115$, but not for $x=0.125$ \cite{nico14}.

It is also relevant to compare with observations of charge-density-wave (CDW) order in \ybco\ \cite{ghir12,chan12a}.  While there are differences in the doping dependence of the charge-ordering wave vectors and the connection to the spin correlations, the CDW order in \ybco\ is strongest for hole concentration near 0.12 \cite{huck14,blan14}, a remarkable similarity to \lbco\ \cite{huck11} and \lnsco\ \cite{ichi00}.  The CDW order in \ybco\ appears at temperatures as high as $\sim150$~K in a lattice with OI order present to above room temperature.

Finally, we note that charge-stripe order has recently been detected by x-ray diffraction in \lsco\ with $x\sim0.12$ \cite{wu12,chri14,crof14,tham14}, confirming an earlier identification by nuclear magnetic resonance \cite{hunt99}.   This result is somewhat surprising, as the average structure of \lsco\ is LTO (although electron diffraction studies have indicated LTT-like regions at LTO twin boundaries for $x=0.12$ \cite{hori97} and LTLO order for $x=0.115$ \cite{koya95}).  A key difference from \lbco\ is that the average orientation of the charge stripes is rotated away from the Cu-O bond direction by a small amount.  Despite this difference, it seems relevant to ask whether dynamic or quasi-static OI tilts may be present in \lsco.  We hope to test this possibility in the future.

%
\section{Summary}
\label{summary}
%

Experimental evidence for the persistence, on a nanometer lengthscale, of \cuosix\ LTT-like octahedral tilt correlations deep into the LTO crystallographic phase in \lbco\ with $0.095 \leq x \leq 0.155$ has been presented. Despite the average structure becoming orthorhombic above $T_{LT}$ as evidenced by Rietveld refinements of neutron scattering data, PDF analysis of the same data shows that the local structure retains its base temperature signatures consistent with orthogonal inequivalent state up to at least $T_{HT}$, where LTO transforms to HTT. The analysis also suggests that there is little preference between OI and OE tilts at high temperature, consistent with the idea that, in the HTT phase, the Cu-O(2) bond precesses rather smoothly around the $c$ axis. The bond-length modulation that pins stripe order is found to be remarkably subtle, with no anomalous bond length disorder at low temperature, placing an upper limit on in-plane Cu-O bondlength anisotropy of 0.005~\AA. Complementary inelastic neutron scattering measurements on $x=1/8$ single crystal sample reveal that upon heating across $T_{LT}$ the tilt correlations become extremely short-range and dynamic. The finite width of excitations around ${\bf Q}=(3,3,L)$ along $Q_z$ further indicates that the LTT tilt fluctuations have 3D character. The relative magnitude of the effect is maximum at $x=1/8$ doping where bulk superconductivity is most strongly suppressed, suggesting that the structural anisotropy might be influenced by electron-phonon coupling and particular stability of the stripe-ordered phase at this composition.

\begin{acknowledgments}
Work at Brookhaven National Laboratory was supported by US DOE, Office of Science, Office of Basic Energy
Sciences (DOE-BES) under contract DE-AC02-98CH10886. Neutron PDF experiments were carried out on NPDF at
LANSCE, funded by DOE BES; Los Alamos National Laboratory is operated by Los Alamos National Security LLC under
contract No. DE-AC52-06NA25396. Inelastic neutron scattering experiments were carried out on HYSPEC at Spallation Neutron
Source at ORNL, sponsored by the Scientific User Facilities Division, Office of Basic Energy Sciences, U.S. Department of Energy.
JMT and RDZ are grateful to B. L. Winn and M. Graves-Brook for assistance with the HYSPEC measurements. ESB gratefully acknowledges T. E. Proffen and J. Siewenie for assistance with the NPDF measurements.
\end{acknowledgments}


\begin{thebibliography}{79}%
\makeatletter
\providecommand \@ifxundefined [1]{%
 \@ifx{#1\undefined}
}%
\providecommand \@ifnum [1]{%
 \ifnum #1\expandafter \@firstoftwo
 \else \expandafter \@secondoftwo
 \fi
}%
\providecommand \@ifx [1]{%
 \ifx #1\expandafter \@firstoftwo
 \else \expandafter \@secondoftwo
 \fi
}%
\providecommand \natexlab [1]{#1}%
\providecommand \enquote  [1]{``#1''}%
\providecommand \bibnamefont  [1]{#1}%
\providecommand \bibfnamefont [1]{#1}%
\providecommand \citenamefont [1]{#1}%
\providecommand \href@noop [0]{\@secondoftwo}%
\providecommand \href [0]{\begingroup \@sanitize@url \@href}%
\providecommand \@href[1]{\@@startlink{#1}\@@href}%
\providecommand \@@href[1]{\endgroup#1\@@endlink}%
\providecommand \@sanitize@url [0]{\catcode `\\12\catcode `\$12\catcode
  `\&12\catcode `\#12\catcode `\^12\catcode `\_12\catcode `\%12\relax}%
\providecommand \@@startlink[1]{}%
\providecommand \@@endlink[0]{}%
\providecommand \url  [0]{\begingroup\@sanitize@url \@url }%
\providecommand \@url [1]{\endgroup\@href {#1}{\urlprefix }}%
\providecommand \urlprefix  [0]{URL }%
\providecommand \Eprint [0]{\href }%
\providecommand \doibase [0]{http://dx.doi.org/}%
\providecommand \selectlanguage [0]{\@gobble}%
\providecommand \bibinfo  [0]{\@secondoftwo}%
\providecommand \bibfield  [0]{\@secondoftwo}%
\providecommand \translation [1]{[#1]}%
\providecommand \BibitemOpen [0]{}%
\providecommand \bibitemStop [0]{}%
\providecommand \bibitemNoStop [0]{.\EOS\space}%
\providecommand \EOS [0]{\spacefactor3000\relax}%
\providecommand \BibitemShut  [1]{\csname bibitem#1\endcsname}%
\let\auto@bib@innerbib\@empty
\bibitem [{\citenamefont {Bednorz}\ and\ \citenamefont
  {M\"uller}(1986)}]{bedn86}%
  \BibitemOpen
  \bibfield  {author} {\bibinfo {author} {\bibfnamefont {J.}~\bibnamefont
  {Bednorz}}\ and\ \bibinfo {author} {\bibfnamefont {K.}~\bibnamefont
  {M\"uller}},\ }\href@noop {} {\bibfield  {journal} {\bibinfo  {journal} {Z.
  Phys. B}\ }\textbf {\bibinfo {volume} {64}},\ \bibinfo {pages} {189}
  (\bibinfo {year} {1986})}\BibitemShut {NoStop}%
\bibitem [{\citenamefont {Jorgensen}\ \emph {et~al.}(1987)\citenamefont
  {Jorgensen}, \citenamefont {Sch\"uttler}, \citenamefont {Hinks},
  \citenamefont {Capone~II}, \citenamefont {Zhang}, \citenamefont {Brodsky},\
  and\ \citenamefont {Scalapino}}]{jorg87}%
  \BibitemOpen
  \bibfield  {author} {\bibinfo {author} {\bibfnamefont {J.~D.}\ \bibnamefont
  {Jorgensen}}, \bibinfo {author} {\bibfnamefont {H.~B.}\ \bibnamefont
  {Sch\"uttler}}, \bibinfo {author} {\bibfnamefont {D.~G.}\ \bibnamefont
  {Hinks}}, \bibinfo {author} {\bibfnamefont {D.~W.}\ \bibnamefont
  {Capone~II}}, \bibinfo {author} {\bibfnamefont {K.}~\bibnamefont {Zhang}},
  \bibinfo {author} {\bibfnamefont {M.~B.}\ \bibnamefont {Brodsky}}, \ and\
  \bibinfo {author} {\bibfnamefont {D.~J.}\ \bibnamefont {Scalapino}},\
  }\href@noop {} {\bibfield  {journal} {\bibinfo  {journal} {Phys. Rev. Lett.}\
  }\textbf {\bibinfo {volume} {58}},\ \bibinfo {pages} {1024} (\bibinfo {year}
  {1987})}\BibitemShut {NoStop}%
\bibitem [{\citenamefont {Paul}\ \emph {et~al.}(1987)\citenamefont {Paul},
  \citenamefont {Balakrishnan}, \citenamefont {Bernhoeft}, \citenamefont
  {David},\ and\ \citenamefont {Harrison}}]{paul87}%
  \BibitemOpen
  \bibfield  {author} {\bibinfo {author} {\bibfnamefont {D.~M.}\ \bibnamefont
  {Paul}}, \bibinfo {author} {\bibfnamefont {G.}~\bibnamefont {Balakrishnan}},
  \bibinfo {author} {\bibfnamefont {N.~R.}\ \bibnamefont {Bernhoeft}}, \bibinfo
  {author} {\bibfnamefont {W.~I.~F.}\ \bibnamefont {David}}, \ and\ \bibinfo
  {author} {\bibfnamefont {W.~T.~A.}\ \bibnamefont {Harrison}},\ }\href@noop {}
  {\bibfield  {journal} {\bibinfo  {journal} {Phys. Rev. Lett.}\ }\textbf
  {\bibinfo {volume} {58}},\ \bibinfo {pages} {1976} (\bibinfo {year}
  {1987})}\BibitemShut {NoStop}%
\bibitem [{\citenamefont {Moodenbaugh}\ \emph {et~al.}(1988)\citenamefont
  {Moodenbaugh}, \citenamefont {Xu}, \citenamefont {Suenaga}, \citenamefont
  {Folkerts},\ and\ \citenamefont {Shelton}}]{mood88}%
  \BibitemOpen
  \bibfield  {author} {\bibinfo {author} {\bibfnamefont {A.~R.}\ \bibnamefont
  {Moodenbaugh}}, \bibinfo {author} {\bibfnamefont {Y.}~\bibnamefont {Xu}},
  \bibinfo {author} {\bibfnamefont {M.}~\bibnamefont {Suenaga}}, \bibinfo
  {author} {\bibfnamefont {T.~J.}\ \bibnamefont {Folkerts}}, \ and\ \bibinfo
  {author} {\bibfnamefont {R.~N.}\ \bibnamefont {Shelton}},\ }\href@noop {}
  {\bibfield  {journal} {\bibinfo  {journal} {Phys. Rev. B}\ }\textbf {\bibinfo
  {volume} {38}},\ \bibinfo {pages} {4596} (\bibinfo {year}
  {1988})}\BibitemShut {NoStop}%
\bibitem [{\citenamefont {Axe}\ \emph {et~al.}(1989)\citenamefont {Axe},
  \citenamefont {Moudden}, \citenamefont {Hohlwein}, \citenamefont {Cox},
  \citenamefont {Mohanty}, \citenamefont {Moodenbaugh},\ and\ \citenamefont
  {Xu}}]{axe89}%
  \BibitemOpen
  \bibfield  {author} {\bibinfo {author} {\bibfnamefont {J.~D.}\ \bibnamefont
  {Axe}}, \bibinfo {author} {\bibfnamefont {A.~H.}\ \bibnamefont {Moudden}},
  \bibinfo {author} {\bibfnamefont {D.}~\bibnamefont {Hohlwein}}, \bibinfo
  {author} {\bibfnamefont {D.~E.}\ \bibnamefont {Cox}}, \bibinfo {author}
  {\bibfnamefont {K.~M.}\ \bibnamefont {Mohanty}}, \bibinfo {author}
  {\bibfnamefont {A.~R.}\ \bibnamefont {Moodenbaugh}}, \ and\ \bibinfo {author}
  {\bibfnamefont {Y.}~\bibnamefont {Xu}},\ }\href@noop {} {\bibfield  {journal}
  {\bibinfo  {journal} {Phys. Rev. Lett.}\ }\textbf {\bibinfo {volume} {62}},\
  \bibinfo {pages} {2751} (\bibinfo {year} {1989})}\BibitemShut {NoStop}%
\bibitem [{\citenamefont {Suzuki}\ and\ \citenamefont
  {Fujita}(1989{\natexlab{a}})}]{suzu89a}%
  \BibitemOpen
  \bibfield  {author} {\bibinfo {author} {\bibfnamefont {T.}~\bibnamefont
  {Suzuki}}\ and\ \bibinfo {author} {\bibfnamefont {T.}~\bibnamefont
  {Fujita}},\ }\href@noop {} {\bibfield  {journal} {\bibinfo  {journal} {J.
  Phys. Soc. Jpn.}\ }\textbf {\bibinfo {volume} {58}},\ \bibinfo {pages} {1883}
  (\bibinfo {year} {1989}{\natexlab{a}})}\BibitemShut {NoStop}%
\bibitem [{\citenamefont {Cox}\ \emph {et~al.}(1989)\citenamefont {Cox},
  \citenamefont {Zolliker}, \citenamefont {Axe}, \citenamefont {Moudden},
  \citenamefont {Moodenbaugh},\ and\ \citenamefont {Xu}}]{cox89}%
  \BibitemOpen
  \bibfield  {author} {\bibinfo {author} {\bibfnamefont {D.~E.}\ \bibnamefont
  {Cox}}, \bibinfo {author} {\bibfnamefont {P.}~\bibnamefont {Zolliker}},
  \bibinfo {author} {\bibfnamefont {J.~D.}\ \bibnamefont {Axe}}, \bibinfo
  {author} {\bibfnamefont {A.~H.}\ \bibnamefont {Moudden}}, \bibinfo {author}
  {\bibfnamefont {A.~R.}\ \bibnamefont {Moodenbaugh}}, \ and\ \bibinfo {author}
  {\bibfnamefont {Y.}~\bibnamefont {Xu}},\ }\href@noop {} {\bibfield  {journal}
  {\bibinfo  {journal} {Mat. Res. Symp. Proc.}\ }\textbf {\bibinfo {volume}
  {156}},\ \bibinfo {pages} {141} (\bibinfo {year} {1989})}\BibitemShut
  {NoStop}%
\bibitem [{\citenamefont {Billinge}\ \emph {et~al.}(1993)\citenamefont
  {Billinge}, \citenamefont {Kwei}, \citenamefont {Lawson}, \citenamefont
  {Thompson},\ and\ \citenamefont {Takagi}}]{bill93}%
  \BibitemOpen
  \bibfield  {author} {\bibinfo {author} {\bibfnamefont {S.~J.~L.}\
  \bibnamefont {Billinge}}, \bibinfo {author} {\bibfnamefont {G.~H.}\
  \bibnamefont {Kwei}}, \bibinfo {author} {\bibfnamefont {A.~C.}\ \bibnamefont
  {Lawson}}, \bibinfo {author} {\bibfnamefont {J.~D.}\ \bibnamefont
  {Thompson}}, \ and\ \bibinfo {author} {\bibfnamefont {H.}~\bibnamefont
  {Takagi}},\ }\href@noop {} {\bibfield  {journal} {\bibinfo  {journal} {Phys.
  Rev. Lett.}\ }\textbf {\bibinfo {volume} {71}},\ \bibinfo {pages} {1903}
  (\bibinfo {year} {1993})}\BibitemShut {NoStop}%
\bibitem [{\citenamefont {Tranquada}\ \emph {et~al.}(1995)\citenamefont
  {Tranquada}, \citenamefont {Sternlieb}, \citenamefont {Axe}, \citenamefont
  {Nakamura},\ and\ \citenamefont {Uchida}}]{tran95a}%
  \BibitemOpen
  \bibfield  {author} {\bibinfo {author} {\bibfnamefont {J.~M.}\ \bibnamefont
  {Tranquada}}, \bibinfo {author} {\bibfnamefont {B.~J.}\ \bibnamefont
  {Sternlieb}}, \bibinfo {author} {\bibfnamefont {J.~D.}\ \bibnamefont {Axe}},
  \bibinfo {author} {\bibfnamefont {Y.}~\bibnamefont {Nakamura}}, \ and\
  \bibinfo {author} {\bibfnamefont {S.}~\bibnamefont {Uchida}},\ }\href@noop {}
  {\bibfield  {journal} {\bibinfo  {journal} {Nature}\ }\textbf {\bibinfo
  {volume} {375}},\ \bibinfo {pages} {561} (\bibinfo {year}
  {1995})}\BibitemShut {NoStop}%
\bibitem [{\citenamefont {Fujita}\ \emph {et~al.}(2004)\citenamefont {Fujita},
  \citenamefont {Goka}, \citenamefont {Yamada}, \citenamefont {Tranquada},\
  and\ \citenamefont {Regnault}}]{fuji04}%
  \BibitemOpen
  \bibfield  {author} {\bibinfo {author} {\bibfnamefont {M.}~\bibnamefont
  {Fujita}}, \bibinfo {author} {\bibfnamefont {H.}~\bibnamefont {Goka}},
  \bibinfo {author} {\bibfnamefont {K.}~\bibnamefont {Yamada}}, \bibinfo
  {author} {\bibfnamefont {J.~M.}\ \bibnamefont {Tranquada}}, \ and\ \bibinfo
  {author} {\bibfnamefont {L.~P.}\ \bibnamefont {Regnault}},\ }\href@noop {}
  {\bibfield  {journal} {\bibinfo  {journal} {Phys. Rev. B}\ }\textbf {\bibinfo
  {volume} {70}},\ \bibinfo {pages} {104517} (\bibinfo {year}
  {2004})}\BibitemShut {NoStop}%
\bibitem [{\citenamefont {Li}\ \emph {et~al.}(2007)\citenamefont {Li},
  \citenamefont {{H\"ucker}}, \citenamefont {Gu}, \citenamefont {Tsvelik},\
  and\ \citenamefont {Tranquada}}]{li07}%
  \BibitemOpen
  \bibfield  {author} {\bibinfo {author} {\bibfnamefont {Q.}~\bibnamefont
  {Li}}, \bibinfo {author} {\bibfnamefont {M.}~\bibnamefont {{H\"ucker}}},
  \bibinfo {author} {\bibfnamefont {G.~D.}\ \bibnamefont {Gu}}, \bibinfo
  {author} {\bibfnamefont {A.~M.}\ \bibnamefont {Tsvelik}}, \ and\ \bibinfo
  {author} {\bibfnamefont {J.~M.}\ \bibnamefont {Tranquada}},\ }\href@noop {}
  {\bibfield  {journal} {\bibinfo  {journal} {Phys. Rev. Lett.}\ }\textbf
  {\bibinfo {volume} {99}},\ \bibinfo {eid} {067001} (\bibinfo {year}
  {2007})}\BibitemShut {NoStop}%
\bibitem [{\citenamefont {Tranquada}\ \emph {et~al.}(2008)\citenamefont
  {Tranquada}, \citenamefont {Gu}, \citenamefont {H{\"u}cker}, \citenamefont
  {Jie}, \citenamefont {Kang}, \citenamefont {Klingeler}, \citenamefont {Li},
  \citenamefont {Tristan}, \citenamefont {Wen}, \citenamefont {Xu},
  \citenamefont {Xu}, \citenamefont {Zhou},\ and\ \citenamefont
  {v.~Zimmermann}}]{tran08}%
  \BibitemOpen
  \bibfield  {author} {\bibinfo {author} {\bibfnamefont {J.~M.}\ \bibnamefont
  {Tranquada}}, \bibinfo {author} {\bibfnamefont {G.~D.}\ \bibnamefont {Gu}},
  \bibinfo {author} {\bibfnamefont {M.}~\bibnamefont {H{\"u}cker}}, \bibinfo
  {author} {\bibfnamefont {Q.}~\bibnamefont {Jie}}, \bibinfo {author}
  {\bibfnamefont {H.-J.}\ \bibnamefont {Kang}}, \bibinfo {author}
  {\bibfnamefont {R.}~\bibnamefont {Klingeler}}, \bibinfo {author}
  {\bibfnamefont {Q.}~\bibnamefont {Li}}, \bibinfo {author} {\bibfnamefont
  {N.}~\bibnamefont {Tristan}}, \bibinfo {author} {\bibfnamefont {J.~S.}\
  \bibnamefont {Wen}}, \bibinfo {author} {\bibfnamefont {G.~Y.}\ \bibnamefont
  {Xu}}, \bibinfo {author} {\bibfnamefont {Z.~J.}\ \bibnamefont {Xu}}, \bibinfo
  {author} {\bibfnamefont {J.}~\bibnamefont {Zhou}}, \ and\ \bibinfo {author}
  {\bibfnamefont {M.}~\bibnamefont {v.~Zimmermann}},\ }\href@noop {} {\bibfield
   {journal} {\bibinfo  {journal} {Phys. Rev. B}\ }\textbf {\bibinfo {volume}
  {78}},\ \bibinfo {eid} {174529} (\bibinfo {year} {2008})}\BibitemShut
  {NoStop}%
\bibitem [{\citenamefont {Berg}\ \emph {et~al.}(2009)\citenamefont {Berg},
  \citenamefont {Fradkin}, \citenamefont {Kivelson},\ and\ \citenamefont
  {Tranquada}}]{berg09b}%
  \BibitemOpen
  \bibfield  {author} {\bibinfo {author} {\bibfnamefont {E.}~\bibnamefont
  {Berg}}, \bibinfo {author} {\bibfnamefont {E.}~\bibnamefont {Fradkin}},
  \bibinfo {author} {\bibfnamefont {S.~A.}\ \bibnamefont {Kivelson}}, \ and\
  \bibinfo {author} {\bibfnamefont {J.~M.}\ \bibnamefont {Tranquada}},\
  }\href@noop {} {\bibfield  {journal} {\bibinfo  {journal} {New J. Phys.}\
  }\textbf {\bibinfo {volume} {11}},\ \bibinfo {pages} {115004} (\bibinfo
  {year} {2009})}\BibitemShut {NoStop}%
\bibitem [{\citenamefont {Himeda}\ \emph {et~al.}(2002)\citenamefont {Himeda},
  \citenamefont {Kato},\ and\ \citenamefont {Ogata}}]{hime02}%
  \BibitemOpen
  \bibfield  {author} {\bibinfo {author} {\bibfnamefont {A.}~\bibnamefont
  {Himeda}}, \bibinfo {author} {\bibfnamefont {T.}~\bibnamefont {Kato}}, \ and\
  \bibinfo {author} {\bibfnamefont {M.}~\bibnamefont {Ogata}},\ }\href@noop {}
  {\bibfield  {journal} {\bibinfo  {journal} {Phys. Rev. Lett.}\ }\textbf
  {\bibinfo {volume} {88}},\ \bibinfo {pages} {117001} (\bibinfo {year}
  {2002})}\BibitemShut {NoStop}%
\bibitem [{\citenamefont {Suzuki}\ and\ \citenamefont
  {Fujita}(1989{\natexlab{b}})}]{suzu89b}%
  \BibitemOpen
  \bibfield  {author} {\bibinfo {author} {\bibfnamefont {T.}~\bibnamefont
  {Suzuki}}\ and\ \bibinfo {author} {\bibfnamefont {T.}~\bibnamefont
  {Fujita}},\ }\href@noop {} {\bibfield  {journal} {\bibinfo  {journal}
  {Physica C}\ }\textbf {\bibinfo {volume} {159}},\ \bibinfo {pages} {111}
  (\bibinfo {year} {1989}{\natexlab{b}})}\BibitemShut {NoStop}%
\bibitem [{\citenamefont {Zhao}\ \emph {et~al.}(2007)\citenamefont {Zhao},
  \citenamefont {Gaulin}, \citenamefont {Castellan}, \citenamefont {Ruff},
  \citenamefont {Dunsiger}, \citenamefont {Gu},\ and\ \citenamefont
  {Dabkowska}}]{zhao07}%
  \BibitemOpen
  \bibfield  {author} {\bibinfo {author} {\bibfnamefont {Y.}~\bibnamefont
  {Zhao}}, \bibinfo {author} {\bibfnamefont {B.~D.}\ \bibnamefont {Gaulin}},
  \bibinfo {author} {\bibfnamefont {J.~P.}\ \bibnamefont {Castellan}}, \bibinfo
  {author} {\bibfnamefont {J.~P.~C.}\ \bibnamefont {Ruff}}, \bibinfo {author}
  {\bibfnamefont {S.~R.}\ \bibnamefont {Dunsiger}}, \bibinfo {author}
  {\bibfnamefont {G.~D.}\ \bibnamefont {Gu}}, \ and\ \bibinfo {author}
  {\bibfnamefont {H.~A.}\ \bibnamefont {Dabkowska}},\ }\href@noop {} {\bibfield
   {journal} {\bibinfo  {journal} {Phys. Rev. B}\ }\textbf {\bibinfo {volume}
  {76}},\ \bibinfo {eid} {184121} (\bibinfo {year} {2007})}\BibitemShut
  {NoStop}%
\bibitem [{\citenamefont {H\"ucker}\ \emph
  {et~al.}(2011{\natexlab{a}})\citenamefont {H\"ucker}, \citenamefont
  {v.~Zimmermann}, \citenamefont {Gu}, \citenamefont {Xu}, \citenamefont {Wen},
  \citenamefont {Xu}, \citenamefont {Kang}, \citenamefont {Zheludev},\ and\
  \citenamefont {Tranquada}}]{huck11}%
  \BibitemOpen
  \bibfield  {author} {\bibinfo {author} {\bibfnamefont {M.}~\bibnamefont
  {H\"ucker}}, \bibinfo {author} {\bibfnamefont {M.}~\bibnamefont
  {v.~Zimmermann}}, \bibinfo {author} {\bibfnamefont {G.~D.}\ \bibnamefont
  {Gu}}, \bibinfo {author} {\bibfnamefont {Z.~J.}\ \bibnamefont {Xu}}, \bibinfo
  {author} {\bibfnamefont {J.~S.}\ \bibnamefont {Wen}}, \bibinfo {author}
  {\bibfnamefont {G.}~\bibnamefont {Xu}}, \bibinfo {author} {\bibfnamefont
  {H.~J.}\ \bibnamefont {Kang}}, \bibinfo {author} {\bibfnamefont
  {A.}~\bibnamefont {Zheludev}}, \ and\ \bibinfo {author} {\bibfnamefont
  {J.~M.}\ \bibnamefont {Tranquada}},\ }\href@noop {} {\bibfield  {journal}
  {\bibinfo  {journal} {Phys. Rev. B}\ }\textbf {\bibinfo {volume} {83}},\
  \bibinfo {pages} {104506} (\bibinfo {year} {2011}{\natexlab{a}})}\BibitemShut
  {NoStop}%
\bibitem [{\citenamefont {Wen}\ \emph {et~al.}(2012)\citenamefont {Wen},
  \citenamefont {Xu}, \citenamefont {Xu}, \citenamefont {Jie}, \citenamefont
  {H\"ucker}, \citenamefont {Zheludev}, \citenamefont {Tian}, \citenamefont
  {Winn}, \citenamefont {Zarestky}, \citenamefont {Singh}, \citenamefont
  {Hong}, \citenamefont {Li}, \citenamefont {Gu},\ and\ \citenamefont
  {Tranquada}}]{wen12a}%
  \BibitemOpen
  \bibfield  {author} {\bibinfo {author} {\bibfnamefont {J.}~\bibnamefont
  {Wen}}, \bibinfo {author} {\bibfnamefont {Z.}~\bibnamefont {Xu}}, \bibinfo
  {author} {\bibfnamefont {G.}~\bibnamefont {Xu}}, \bibinfo {author}
  {\bibfnamefont {Q.}~\bibnamefont {Jie}}, \bibinfo {author} {\bibfnamefont
  {M.}~\bibnamefont {H\"ucker}}, \bibinfo {author} {\bibfnamefont
  {A.}~\bibnamefont {Zheludev}}, \bibinfo {author} {\bibfnamefont
  {W.}~\bibnamefont {Tian}}, \bibinfo {author} {\bibfnamefont {B.~L.}\
  \bibnamefont {Winn}}, \bibinfo {author} {\bibfnamefont {J.~L.}\ \bibnamefont
  {Zarestky}}, \bibinfo {author} {\bibfnamefont {D.~K.}\ \bibnamefont {Singh}},
  \bibinfo {author} {\bibfnamefont {T.}~\bibnamefont {Hong}}, \bibinfo {author}
  {\bibfnamefont {Q.}~\bibnamefont {Li}}, \bibinfo {author} {\bibfnamefont
  {G.}~\bibnamefont {Gu}}, \ and\ \bibinfo {author} {\bibfnamefont {J.~M.}\
  \bibnamefont {Tranquada}},\ }\href@noop {} {\bibfield  {journal} {\bibinfo
  {journal} {Phys. Rev. B}\ }\textbf {\bibinfo {volume} {85}},\ \bibinfo
  {pages} {134512} (\bibinfo {year} {2012})}\BibitemShut {NoStop}%
\bibitem [{\citenamefont {Billinge}\ \emph
  {et~al.}(1994{\natexlab{a}})\citenamefont {Billinge}, \citenamefont {Kwei},\
  and\ \citenamefont {Takagi}}]{bill94}%
  \BibitemOpen
  \bibfield  {author} {\bibinfo {author} {\bibfnamefont {S.~J.~L.}\
  \bibnamefont {Billinge}}, \bibinfo {author} {\bibfnamefont {G.~H.}\
  \bibnamefont {Kwei}}, \ and\ \bibinfo {author} {\bibfnamefont
  {H.}~\bibnamefont {Takagi}},\ }\href@noop {} {\bibfield  {journal} {\bibinfo
  {journal} {Phys. Rev. Lett.}\ }\textbf {\bibinfo {volume} {72}},\ \bibinfo
  {pages} {2282} (\bibinfo {year} {1994}{\natexlab{a}})}\BibitemShut {NoStop}%
\bibitem [{\citenamefont {Haskel}\ \emph {et~al.}(2000)\citenamefont {Haskel},
  \citenamefont {Stern}, \citenamefont {Dogan},\ and\ \citenamefont
  {Moodenbaugh}}]{hask00}%
  \BibitemOpen
  \bibfield  {author} {\bibinfo {author} {\bibfnamefont {D.}~\bibnamefont
  {Haskel}}, \bibinfo {author} {\bibfnamefont {E.~A.}\ \bibnamefont {Stern}},
  \bibinfo {author} {\bibfnamefont {F.}~\bibnamefont {Dogan}}, \ and\ \bibinfo
  {author} {\bibfnamefont {A.~R.}\ \bibnamefont {Moodenbaugh}},\ }\href@noop {}
  {\bibfield  {journal} {\bibinfo  {journal} {Phys. Rev. B}\ }\textbf {\bibinfo
  {volume} {61}},\ \bibinfo {pages} {7055} (\bibinfo {year}
  {2000})}\BibitemShut {NoStop}%
\bibitem [{\citenamefont {Bo\v{z}in}\ \emph {et~al.}(1997)\citenamefont
  {Bo\v{z}in}, \citenamefont {Billinge},\ and\ \citenamefont {Kwei}}]{bozi97}%
  \BibitemOpen
  \bibfield  {author} {\bibinfo {author} {\bibfnamefont {E.}~\bibnamefont
  {Bo\v{z}in}}, \bibinfo {author} {\bibfnamefont {S.}~\bibnamefont {Billinge}},
  \ and\ \bibinfo {author} {\bibfnamefont {G.}~\bibnamefont {Kwei}},\
  }\href@noop {} {\bibfield  {journal} {\bibinfo  {journal} {Physica B}\
  }\textbf {\bibinfo {volume} {241---243}},\ \bibinfo {pages} {795} (\bibinfo
  {year} {1997})}\BibitemShut {NoStop}%
\bibitem [{\citenamefont {Fabbris}\ \emph {et~al.}(2013)\citenamefont
  {Fabbris}, \citenamefont {H\"ucker}, \citenamefont {Gu}, \citenamefont
  {Tranquada},\ and\ \citenamefont {Haskel}}]{fabb13}%
  \BibitemOpen
  \bibfield  {author} {\bibinfo {author} {\bibfnamefont {G.}~\bibnamefont
  {Fabbris}}, \bibinfo {author} {\bibfnamefont {M.}~\bibnamefont {H\"ucker}},
  \bibinfo {author} {\bibfnamefont {G.~D.}\ \bibnamefont {Gu}}, \bibinfo
  {author} {\bibfnamefont {J.~M.}\ \bibnamefont {Tranquada}}, \ and\ \bibinfo
  {author} {\bibfnamefont {D.}~\bibnamefont {Haskel}},\ }\href@noop {}
  {\bibfield  {journal} {\bibinfo  {journal} {Phys. Rev. B}\ }\textbf {\bibinfo
  {volume} {88}},\ \bibinfo {pages} {060507} (\bibinfo {year}
  {2013})}\BibitemShut {NoStop}%
\bibitem [{\citenamefont {Axe}\ and\ \citenamefont {Crawford}(1994)}]{axe94}%
  \BibitemOpen
  \bibfield  {author} {\bibinfo {author} {\bibfnamefont {J.~D.}\ \bibnamefont
  {Axe}}\ and\ \bibinfo {author} {\bibfnamefont {M.~K.}\ \bibnamefont
  {Crawford}},\ }\href@noop {} {\bibfield  {journal} {\bibinfo  {journal} {J.
  Low Temp. Phys.}\ }\textbf {\bibinfo {volume} {95}},\ \bibinfo {pages} {271}
  (\bibinfo {year} {1994})}\BibitemShut {NoStop}%
\bibitem [{\citenamefont {Radaelli}\ \emph {et~al.}(1994)\citenamefont
  {Radaelli}, \citenamefont {Hinks}, \citenamefont {Mitchell}, \citenamefont
  {Hunter}, \citenamefont {Wagner}, \citenamefont {Dabrowski}, \citenamefont
  {Vandervoort}, \citenamefont {Viswanathan},\ and\ \citenamefont
  {Jorgensen}}]{rada94}%
  \BibitemOpen
  \bibfield  {author} {\bibinfo {author} {\bibfnamefont {P.~G.}\ \bibnamefont
  {Radaelli}}, \bibinfo {author} {\bibfnamefont {D.~G.}\ \bibnamefont {Hinks}},
  \bibinfo {author} {\bibfnamefont {A.~W.}\ \bibnamefont {Mitchell}}, \bibinfo
  {author} {\bibfnamefont {B.~A.}\ \bibnamefont {Hunter}}, \bibinfo {author}
  {\bibfnamefont {J.~L.}\ \bibnamefont {Wagner}}, \bibinfo {author}
  {\bibfnamefont {B.}~\bibnamefont {Dabrowski}}, \bibinfo {author}
  {\bibfnamefont {K.~G.}\ \bibnamefont {Vandervoort}}, \bibinfo {author}
  {\bibfnamefont {H.~K.}\ \bibnamefont {Viswanathan}}, \ and\ \bibinfo {author}
  {\bibfnamefont {J.~D.}\ \bibnamefont {Jorgensen}},\ }\href@noop {} {\bibfield
   {journal} {\bibinfo  {journal} {Phys. Rev. B}\ }\textbf {\bibinfo {volume}
  {49}},\ \bibinfo {pages} {4163} (\bibinfo {year} {1994})}\BibitemShut
  {NoStop}%
\bibitem [{\citenamefont {H\"ucker}\ \emph
  {et~al.}(2011{\natexlab{b}})\citenamefont {H\"ucker}, \citenamefont
  {v.~Zimmermann}, \citenamefont {Gu}, \citenamefont {Xu}, \citenamefont {Wen},
  \citenamefont {Xu}, \citenamefont {Kang}, \citenamefont {Zheludev},\ and\
  \citenamefont {Tranquada}}]{hucke;prb11}%
  \BibitemOpen
  \bibfield  {author} {\bibinfo {author} {\bibfnamefont {M.}~\bibnamefont
  {H\"ucker}}, \bibinfo {author} {\bibfnamefont {M.}~\bibnamefont
  {v.~Zimmermann}}, \bibinfo {author} {\bibfnamefont {G.~D.}\ \bibnamefont
  {Gu}}, \bibinfo {author} {\bibfnamefont {Z.~J.}\ \bibnamefont {Xu}}, \bibinfo
  {author} {\bibfnamefont {J.~S.}\ \bibnamefont {Wen}}, \bibinfo {author}
  {\bibfnamefont {G.}~\bibnamefont {Xu}}, \bibinfo {author} {\bibfnamefont
  {H.~J.}\ \bibnamefont {Kang}}, \bibinfo {author} {\bibfnamefont
  {A.}~\bibnamefont {Zheludev}}, \ and\ \bibinfo {author} {\bibfnamefont
  {J.~M.}\ \bibnamefont {Tranquada}},\ }\href {\doibase
  10.1103/PhysRevB.83.104506} {\bibfield  {journal} {\bibinfo  {journal} {Phys.
  Rev. B}\ }\textbf {\bibinfo {volume} {83}},\ \bibinfo {pages} {104506}
  (\bibinfo {year} {2011}{\natexlab{b}})}\BibitemShut {NoStop}%
\bibitem [{\citenamefont {Crawford}\ \emph {et~al.}(1991)\citenamefont
  {Crawford}, \citenamefont {Harlow}, \citenamefont {McCarron}, \citenamefont
  {Farneth}, \citenamefont {Axe}, \citenamefont {Chou},\ and\ \citenamefont
  {Huang}}]{craw91}%
  \BibitemOpen
  \bibfield  {author} {\bibinfo {author} {\bibfnamefont {M.~K.}\ \bibnamefont
  {Crawford}}, \bibinfo {author} {\bibfnamefont {R.~L.}\ \bibnamefont
  {Harlow}}, \bibinfo {author} {\bibfnamefont {E.~M.}\ \bibnamefont
  {McCarron}}, \bibinfo {author} {\bibfnamefont {W.~E.}\ \bibnamefont
  {Farneth}}, \bibinfo {author} {\bibfnamefont {J.~D.}\ \bibnamefont {Axe}},
  \bibinfo {author} {\bibfnamefont {H.}~\bibnamefont {Chou}}, \ and\ \bibinfo
  {author} {\bibfnamefont {Q.}~\bibnamefont {Huang}},\ }\href@noop {}
  {\bibfield  {journal} {\bibinfo  {journal} {Phys. Rev. B}\ }\textbf {\bibinfo
  {volume} {44}},\ \bibinfo {pages} {7749} (\bibinfo {year}
  {1991})}\BibitemShut {NoStop}%
\bibitem [{\citenamefont {Birgeneau}\ \emph {et~al.}(1987)\citenamefont
  {Birgeneau}, \citenamefont {Chen}, \citenamefont {Gabbe}, \citenamefont
  {Jenssen}, \citenamefont {Kastner}, \citenamefont {Peters}, \citenamefont
  {Picone}, \citenamefont {Thio}, \citenamefont {Thurston}, \citenamefont
  {Tuller}, \citenamefont {Axe}, \citenamefont {B\"oni},\ and\ \citenamefont
  {Shirane}}]{birg87}%
  \BibitemOpen
  \bibfield  {author} {\bibinfo {author} {\bibfnamefont {R.~J.}\ \bibnamefont
  {Birgeneau}}, \bibinfo {author} {\bibfnamefont {C.~Y.}\ \bibnamefont {Chen}},
  \bibinfo {author} {\bibfnamefont {D.~R.}\ \bibnamefont {Gabbe}}, \bibinfo
  {author} {\bibfnamefont {H.~P.}\ \bibnamefont {Jenssen}}, \bibinfo {author}
  {\bibfnamefont {M.~A.}\ \bibnamefont {Kastner}}, \bibinfo {author}
  {\bibfnamefont {C.~J.}\ \bibnamefont {Peters}}, \bibinfo {author}
  {\bibfnamefont {P.~J.}\ \bibnamefont {Picone}}, \bibinfo {author}
  {\bibfnamefont {T.}~\bibnamefont {Thio}}, \bibinfo {author} {\bibfnamefont
  {T.~R.}\ \bibnamefont {Thurston}}, \bibinfo {author} {\bibfnamefont {H.~L.}\
  \bibnamefont {Tuller}}, \bibinfo {author} {\bibfnamefont {J.~D.}\
  \bibnamefont {Axe}}, \bibinfo {author} {\bibfnamefont {P.}~\bibnamefont
  {B\"oni}}, \ and\ \bibinfo {author} {\bibfnamefont {G.}~\bibnamefont
  {Shirane}},\ }\href@noop {} {\bibfield  {journal} {\bibinfo  {journal} {Phys.
  Rev. Lett.}\ }\textbf {\bibinfo {volume} {59}},\ \bibinfo {pages} {1329}
  (\bibinfo {year} {1987})}\BibitemShut {NoStop}%
\bibitem [{\citenamefont {Thurston}\ \emph {et~al.}(1989)\citenamefont
  {Thurston}, \citenamefont {Birgeneau}, \citenamefont {Gabbe}, \citenamefont
  {Jenssen}, \citenamefont {Kastner}, \citenamefont {Picone}, \citenamefont
  {Preyer}, \citenamefont {Axe}, \citenamefont {B\"oni}, \citenamefont
  {Shirane}, \citenamefont {Sato}, \citenamefont {Fukuda},\ and\ \citenamefont
  {Shamoto}}]{thur89a}%
  \BibitemOpen
  \bibfield  {author} {\bibinfo {author} {\bibfnamefont {T.~R.}\ \bibnamefont
  {Thurston}}, \bibinfo {author} {\bibfnamefont {R.~J.}\ \bibnamefont
  {Birgeneau}}, \bibinfo {author} {\bibfnamefont {D.~R.}\ \bibnamefont
  {Gabbe}}, \bibinfo {author} {\bibfnamefont {H.~P.}\ \bibnamefont {Jenssen}},
  \bibinfo {author} {\bibfnamefont {M.~A.}\ \bibnamefont {Kastner}}, \bibinfo
  {author} {\bibfnamefont {P.~J.}\ \bibnamefont {Picone}}, \bibinfo {author}
  {\bibfnamefont {N.~W.}\ \bibnamefont {Preyer}}, \bibinfo {author}
  {\bibfnamefont {J.~D.}\ \bibnamefont {Axe}}, \bibinfo {author} {\bibfnamefont
  {P.}~\bibnamefont {B\"oni}}, \bibinfo {author} {\bibfnamefont
  {G.}~\bibnamefont {Shirane}}, \bibinfo {author} {\bibfnamefont
  {M.}~\bibnamefont {Sato}}, \bibinfo {author} {\bibfnamefont {K.}~\bibnamefont
  {Fukuda}}, \ and\ \bibinfo {author} {\bibfnamefont {S.}~\bibnamefont
  {Shamoto}},\ }\href@noop {} {\bibfield  {journal} {\bibinfo  {journal} {Phys.
  Rev. B}\ }\textbf {\bibinfo {volume} {39}},\ \bibinfo {pages} {4327}
  (\bibinfo {year} {1989})}\BibitemShut {NoStop}%
\bibitem [{\citenamefont {Kimura}\ \emph {et~al.}(2000)\citenamefont {Kimura},
  \citenamefont {Hirota}, \citenamefont {Lee}, \citenamefont {Yamada},\ and\
  \citenamefont {Shirane}}]{kimu00}%
  \BibitemOpen
  \bibfield  {author} {\bibinfo {author} {\bibfnamefont {H.}~\bibnamefont
  {Kimura}}, \bibinfo {author} {\bibfnamefont {K.}~\bibnamefont {Hirota}},
  \bibinfo {author} {\bibfnamefont {C.-H.}\ \bibnamefont {Lee}}, \bibinfo
  {author} {\bibfnamefont {K.}~\bibnamefont {Yamada}}, \ and\ \bibinfo {author}
  {\bibfnamefont {G.}~\bibnamefont {Shirane}},\ }\href@noop {} {\bibfield
  {journal} {\bibinfo  {journal} {J. Phys. Soc. Jpn.}\ }\textbf {\bibinfo
  {volume} {69}},\ \bibinfo {pages} {851} (\bibinfo {year} {2000})}\BibitemShut
  {NoStop}%
\bibitem [{\citenamefont {Keimer}\ \emph {et~al.}(1993)\citenamefont {Keimer},
  \citenamefont {Birgeneau}, \citenamefont {Cassanho}, \citenamefont {Endoh},
  \citenamefont {Greven}, \citenamefont {Kastner},\ and\ \citenamefont
  {Shirane}}]{keim93}%
  \BibitemOpen
  \bibfield  {author} {\bibinfo {author} {\bibfnamefont {B.}~\bibnamefont
  {Keimer}}, \bibinfo {author} {\bibfnamefont {R.~J.}\ \bibnamefont
  {Birgeneau}}, \bibinfo {author} {\bibfnamefont {A.}~\bibnamefont {Cassanho}},
  \bibinfo {author} {\bibfnamefont {Y.}~\bibnamefont {Endoh}}, \bibinfo
  {author} {\bibfnamefont {M.}~\bibnamefont {Greven}}, \bibinfo {author}
  {\bibfnamefont {M.~A.}\ \bibnamefont {Kastner}}, \ and\ \bibinfo {author}
  {\bibfnamefont {G.}~\bibnamefont {Shirane}},\ }\href@noop {} {\bibfield
  {journal} {\bibinfo  {journal} {Z. Phys. B}\ }\textbf {\bibinfo {volume}
  {91}},\ \bibinfo {pages} {373} (\bibinfo {year} {1993})}\BibitemShut
  {NoStop}%
\bibitem [{\citenamefont {Kimura}\ \emph {et~al.}(2005)\citenamefont {Kimura},
  \citenamefont {Noda}, \citenamefont {Goka}, \citenamefont {Fujita},
  \citenamefont {Yamada},\ and\ \citenamefont {Shirane}}]{kimu05}%
  \BibitemOpen
  \bibfield  {author} {\bibinfo {author} {\bibfnamefont {H.}~\bibnamefont
  {Kimura}}, \bibinfo {author} {\bibfnamefont {Y.}~\bibnamefont {Noda}},
  \bibinfo {author} {\bibfnamefont {H.}~\bibnamefont {Goka}}, \bibinfo {author}
  {\bibfnamefont {M.}~\bibnamefont {Fujita}}, \bibinfo {author} {\bibfnamefont
  {K.}~\bibnamefont {Yamada}}, \ and\ \bibinfo {author} {\bibfnamefont
  {G.}~\bibnamefont {Shirane}},\ }\href@noop {} {\bibfield  {journal} {\bibinfo
   {journal} {J. Phys. Soc. Jpn.}\ }\textbf {\bibinfo {volume} {74}},\ \bibinfo
  {pages} {445} (\bibinfo {year} {2005})}\BibitemShut {NoStop}%
\bibitem [{\citenamefont {Wakimoto}\ \emph {et~al.}(2006)\citenamefont
  {Wakimoto}, \citenamefont {Kimura}, \citenamefont {Fujita}, \citenamefont
  {Yamada}, \citenamefont {Noda}, \citenamefont {Shirane}, \citenamefont {Gu},
  \citenamefont {Kim},\ and\ \citenamefont {Birgeneau}}]{waki06}%
  \BibitemOpen
  \bibfield  {author} {\bibinfo {author} {\bibfnamefont {S.}~\bibnamefont
  {Wakimoto}}, \bibinfo {author} {\bibfnamefont {H.}~\bibnamefont {Kimura}},
  \bibinfo {author} {\bibfnamefont {M.}~\bibnamefont {Fujita}}, \bibinfo
  {author} {\bibfnamefont {K.}~\bibnamefont {Yamada}}, \bibinfo {author}
  {\bibfnamefont {Y.}~\bibnamefont {Noda}}, \bibinfo {author} {\bibfnamefont
  {G.}~\bibnamefont {Shirane}}, \bibinfo {author} {\bibfnamefont
  {G.}~\bibnamefont {Gu}}, \bibinfo {author} {\bibfnamefont {H.}~\bibnamefont
  {Kim}}, \ and\ \bibinfo {author} {\bibfnamefont {R.~J.}\ \bibnamefont
  {Birgeneau}},\ }\href@noop {} {\bibfield  {journal} {\bibinfo  {journal} {J.
  Phys. Soc. Jpn.}\ }\textbf {\bibinfo {volume} {75}},\ \bibinfo {pages}
  {074714} (\bibinfo {year} {2006})}\BibitemShut {NoStop}%
\bibitem [{\citenamefont {Pickett}\ \emph {et~al.}(1991)\citenamefont
  {Pickett}, \citenamefont {Cohen},\ and\ \citenamefont {Krakauer}}]{pick91}%
  \BibitemOpen
  \bibfield  {author} {\bibinfo {author} {\bibfnamefont {W.~E.}\ \bibnamefont
  {Pickett}}, \bibinfo {author} {\bibfnamefont {R.~E.}\ \bibnamefont {Cohen}},
  \ and\ \bibinfo {author} {\bibfnamefont {H.}~\bibnamefont {Krakauer}},\
  }\href@noop {} {\bibfield  {journal} {\bibinfo  {journal} {prl}\ }\textbf
  {\bibinfo {volume} {67}},\ \bibinfo {pages} {228} (\bibinfo {year}
  {1991})}\BibitemShut {NoStop}%
\bibitem [{\citenamefont {Cai}\ and\ \citenamefont {Welch}(1994)}]{cai94}%
  \BibitemOpen
  \bibfield  {author} {\bibinfo {author} {\bibfnamefont {Z.-X.}\ \bibnamefont
  {Cai}}\ and\ \bibinfo {author} {\bibfnamefont {D.~O.}\ \bibnamefont
  {Welch}},\ }\href@noop {} {\bibfield  {journal} {\bibinfo  {journal} {Physica
  C}\ }\textbf {\bibinfo {volume} {231}},\ \bibinfo {pages} {383} (\bibinfo
  {year} {1994})}\BibitemShut {NoStop}%
\bibitem [{\citenamefont {Egami}\ and\ \citenamefont
  {Billinge}(2012)}]{egami;b;utbp12}%
  \BibitemOpen
  \bibfield  {author} {\bibinfo {author} {\bibfnamefont {T.}~\bibnamefont
  {Egami}}\ and\ \bibinfo {author} {\bibfnamefont {S.~J.~L.}\ \bibnamefont
  {Billinge}},\ }\href@noop {} {\emph {\bibinfo {title} {Underneath the Bragg
  peaks: structural analysis of complex materials}}},\ \bibinfo {edition}
  {2nd}\ ed.\ (\bibinfo  {publisher} {Elsevier},\ \bibinfo {address}
  {Amsterdam},\ \bibinfo {year} {2012})\BibitemShut {NoStop}%
\bibitem [{\citenamefont {Peterson}\ \emph {et~al.}(2000)\citenamefont
  {Peterson}, \citenamefont {Gutmann}, \citenamefont {Proffen},\ and\
  \citenamefont {Billinge}}]{peter;jac00}%
  \BibitemOpen
  \bibfield  {author} {\bibinfo {author} {\bibfnamefont {P.~F.}\ \bibnamefont
  {Peterson}}, \bibinfo {author} {\bibfnamefont {M.}~\bibnamefont {Gutmann}},
  \bibinfo {author} {\bibfnamefont {T.}~\bibnamefont {Proffen}}, \ and\
  \bibinfo {author} {\bibfnamefont {S.~J.~L.}\ \bibnamefont {Billinge}},\
  }\href@noop {} {\bibfield  {journal} {\bibinfo  {journal} {J. Appl.
  Crystallogr.}\ }\textbf {\bibinfo {volume} {33}},\ \bibinfo {pages} {1192}
  (\bibinfo {year} {2000})}\BibitemShut {NoStop}%
\bibitem [{\citenamefont {Rietveld}(1967)}]{rietv;ac67}%
  \BibitemOpen
  \bibfield  {author} {\bibinfo {author} {\bibfnamefont {H.~M.}\ \bibnamefont
  {Rietveld}},\ }\href@noop {} {\bibfield  {journal} {\bibinfo  {journal} {Acta
  Crystallogr.}\ }\textbf {\bibinfo {volume} {22}},\ \bibinfo {pages} {151 }
  (\bibinfo {year} {1967})}\BibitemShut {NoStop}%
\bibitem [{\citenamefont {Larson}\ and\ \citenamefont {{Von
  Dreele}}(1987)}]{larso;unpub87}%
  \BibitemOpen
  \bibfield  {author} {\bibinfo {author} {\bibfnamefont {A.~C.}\ \bibnamefont
  {Larson}}\ and\ \bibinfo {author} {\bibfnamefont {R.~B.}\ \bibnamefont {{Von
  Dreele}}},\ }\href@noop {} {\enquote {\bibinfo {title} {General structure
  analysis system},}\ } (\bibinfo {year} {1987}),\ \bibinfo {note} {report No.
  LAUR-86-748, Los Alamos National Laboratory, Los Alamos, NM
  87545}\BibitemShut {NoStop}%
\bibitem [{\citenamefont {Toby}(2001)}]{toby;jac01}%
  \BibitemOpen
  \bibfield  {author} {\bibinfo {author} {\bibfnamefont {B.~H.}\ \bibnamefont
  {Toby}},\ }\href {\doibase 10.1107/S0021889801002242} {\bibfield  {journal}
  {\bibinfo  {journal} {J. Appl. Crystallogr.}\ }\textbf {\bibinfo {volume}
  {34}},\ \bibinfo {pages} {201} (\bibinfo {year} {2001})}\BibitemShut
  {NoStop}%
\bibitem [{\citenamefont {Farrow}\ \emph {et~al.}(2007)\citenamefont {Farrow},
  \citenamefont {Juh\'as}, \citenamefont {Liu}, \citenamefont {Bryndin},
  \citenamefont {{Bo\v zin}}, \citenamefont {Bloch}, \citenamefont {Proffen},\
  and\ \citenamefont {Billinge}}]{farro;jpcm07}%
  \BibitemOpen
  \bibfield  {author} {\bibinfo {author} {\bibfnamefont {C.~L.}\ \bibnamefont
  {Farrow}}, \bibinfo {author} {\bibfnamefont {P.}~\bibnamefont {Juh\'as}},
  \bibinfo {author} {\bibfnamefont {J.}~\bibnamefont {Liu}}, \bibinfo {author}
  {\bibfnamefont {D.}~\bibnamefont {Bryndin}}, \bibinfo {author} {\bibfnamefont
  {E.~S.}\ \bibnamefont {{Bo\v zin}}}, \bibinfo {author} {\bibfnamefont
  {J.}~\bibnamefont {Bloch}}, \bibinfo {author} {\bibfnamefont
  {T.}~\bibnamefont {Proffen}}, \ and\ \bibinfo {author} {\bibfnamefont
  {S.~J.~L.}\ \bibnamefont {Billinge}},\ }\href {\doibase
  10.1088/0953-8984/19/33/335219} {\bibfield  {journal} {\bibinfo  {journal}
  {J. Phys: Condens. Mat.}\ }\textbf {\bibinfo {volume} {19}},\ \bibinfo
  {pages} {335219} (\bibinfo {year} {2007})}\BibitemShut {NoStop}%
\bibitem [{\citenamefont {Winn}\ \emph {et~al.}(2014)\citenamefont {Winn},
  \citenamefont {Filges}, \citenamefont {Garlea}, \citenamefont {Graves-Brook},
  \citenamefont {Hagen}, \citenamefont {Jiang}, \citenamefont {Kenzelmann},
  \citenamefont {Passell}, \citenamefont {Shapiro}, \citenamefont {Tong},\ and\
  \citenamefont {Zaliznyak}}]{winn14}%
  \BibitemOpen
  \bibfield  {author} {\bibinfo {author} {\bibfnamefont {B.}~\bibnamefont
  {Winn}}, \bibinfo {author} {\bibfnamefont {U.}~\bibnamefont {Filges}},
  \bibinfo {author} {\bibfnamefont {V.~O.}\ \bibnamefont {Garlea}}, \bibinfo
  {author} {\bibfnamefont {M.}~\bibnamefont {Graves-Brook}}, \bibinfo {author}
  {\bibfnamefont {M.}~\bibnamefont {Hagen}}, \bibinfo {author} {\bibfnamefont
  {C.}~\bibnamefont {Jiang}}, \bibinfo {author} {\bibfnamefont
  {M.}~\bibnamefont {Kenzelmann}}, \bibinfo {author} {\bibfnamefont
  {L.}~\bibnamefont {Passell}}, \bibinfo {author} {\bibfnamefont {S.~M.}\
  \bibnamefont {Shapiro}}, \bibinfo {author} {\bibfnamefont {X.}~\bibnamefont
  {Tong}}, \ and\ \bibinfo {author} {\bibfnamefont {I.}~\bibnamefont
  {Zaliznyak}},\ }\href@noop {} {\bibfield  {journal} {\bibinfo  {journal} {EPJ
  Web of Conferences}\ } (\bibinfo {year} {2014})}\BibitemShut {NoStop}%
\bibitem [{\citenamefont {Arnold}\ \emph {et~al.}(2014)\citenamefont {Arnold},
  \citenamefont {Bilheux}, \citenamefont {Borreguero}, \citenamefont {Butsa},
  \citenamefont {Campbell}, \citenamefont {Chapona}, \citenamefont {Doucet},
  \citenamefont {Draper}, \citenamefont {{Ferraz Leal}}, \citenamefont {Gigga},
  \citenamefont {Lynch}, \citenamefont {Markvardsen}, \citenamefont
  {Mikkelsone}, \citenamefont {Mikkelsone}, \citenamefont {Miller},
  \citenamefont {Palmen}, \citenamefont {Parker}, \citenamefont {Passos},
  \citenamefont {Perring}, \citenamefont {Peterson}, \citenamefont {Ren},
  \citenamefont {Reuter}, \citenamefont {Savici}, \citenamefont {Taylor},
  \citenamefont {Taylor}, \citenamefont {Tolchenov}, \citenamefont {Zhou},\
  and\ \citenamefont {Zikovsky}}]{arnold14}%
  \BibitemOpen
  \bibfield  {author} {\bibinfo {author} {\bibfnamefont {O.}~\bibnamefont
  {Arnold}}, \bibinfo {author} {\bibfnamefont {J.~C.}\ \bibnamefont {Bilheux}},
  \bibinfo {author} {\bibfnamefont {J.~M.}\ \bibnamefont {Borreguero}},
  \bibinfo {author} {\bibfnamefont {A.}~\bibnamefont {Butsa}}, \bibinfo
  {author} {\bibfnamefont {S.~I.}\ \bibnamefont {Campbell}}, \bibinfo {author}
  {\bibfnamefont {L.}~\bibnamefont {Chapona}}, \bibinfo {author} {\bibfnamefont
  {M.}~\bibnamefont {Doucet}}, \bibinfo {author} {\bibfnamefont
  {N.}~\bibnamefont {Draper}}, \bibinfo {author} {\bibfnamefont
  {R.}~\bibnamefont {{Ferraz Leal}}}, \bibinfo {author} {\bibfnamefont {M.~A.}\
  \bibnamefont {Gigga}}, \bibinfo {author} {\bibfnamefont {V.~E.}\ \bibnamefont
  {Lynch}}, \bibinfo {author} {\bibfnamefont {A.}~\bibnamefont {Markvardsen}},
  \bibinfo {author} {\bibfnamefont {D.~J.}\ \bibnamefont {Mikkelsone}},
  \bibinfo {author} {\bibfnamefont {R.~L.}\ \bibnamefont {Mikkelsone}},
  \bibinfo {author} {\bibfnamefont {R.}~\bibnamefont {Miller}}, \bibinfo
  {author} {\bibfnamefont {K.}~\bibnamefont {Palmen}}, \bibinfo {author}
  {\bibfnamefont {P.}~\bibnamefont {Parker}}, \bibinfo {author} {\bibfnamefont
  {G.}~\bibnamefont {Passos}}, \bibinfo {author} {\bibfnamefont {T.~G.}\
  \bibnamefont {Perring}}, \bibinfo {author} {\bibfnamefont {P.~F.}\
  \bibnamefont {Peterson}}, \bibinfo {author} {\bibfnamefont {S.}~\bibnamefont
  {Ren}}, \bibinfo {author} {\bibfnamefont {M.~A.}\ \bibnamefont {Reuter}},
  \bibinfo {author} {\bibfnamefont {A.~T.}\ \bibnamefont {Savici}}, \bibinfo
  {author} {\bibfnamefont {J.~W.}\ \bibnamefont {Taylor}}, \bibinfo {author}
  {\bibfnamefont {R.~J.}\ \bibnamefont {Taylor}}, \bibinfo {author}
  {\bibfnamefont {R.}~\bibnamefont {Tolchenov}}, \bibinfo {author}
  {\bibfnamefont {W.}~\bibnamefont {Zhou}}, \ and\ \bibinfo {author}
  {\bibfnamefont {J.}~\bibnamefont {Zikovsky}},\ }\href@noop {} {\bibfield
  {journal} {\bibinfo  {journal} {Nucl. Instrum. Meth. A}\ }\textbf {\bibinfo
  {volume} {764}},\ \bibinfo {pages} {156–166} (\bibinfo {year}
  {2014})}\BibitemShut {NoStop}%
\bibitem [{\citenamefont {Azuah}\ \emph {et~al.}(2009)\citenamefont {Azuah},
  \citenamefont {Kneller}, \citenamefont {Qiu}, \citenamefont
  {Tregenna-Piggott}, \citenamefont {Brown}, \citenamefont {Copley},\ and\
  \citenamefont {Dimeo}}]{dave09}%
  \BibitemOpen
  \bibfield  {author} {\bibinfo {author} {\bibfnamefont {R.~T.}\ \bibnamefont
  {Azuah}}, \bibinfo {author} {\bibfnamefont {L.~R.}\ \bibnamefont {Kneller}},
  \bibinfo {author} {\bibfnamefont {Y.}~\bibnamefont {Qiu}}, \bibinfo {author}
  {\bibfnamefont {P.~L.~W.}\ \bibnamefont {Tregenna-Piggott}}, \bibinfo
  {author} {\bibfnamefont {C.~M.}\ \bibnamefont {Brown}}, \bibinfo {author}
  {\bibfnamefont {J.~R.~D.}\ \bibnamefont {Copley}}, \ and\ \bibinfo {author}
  {\bibfnamefont {R.~M.}\ \bibnamefont {Dimeo}},\ }\href {\doibase
  http://dx.doi.org/10.6028/jres.114.025} {\bibfield  {journal} {\bibinfo
  {journal} {J. Res. Natl. Inst. Stan. Technol.}\ }\textbf {\bibinfo {volume}
  {114}},\ \bibinfo {pages} {341} (\bibinfo {year} {2009})}\BibitemShut
  {NoStop}%
\bibitem [{\citenamefont {Qiu}\ \emph {et~al.}(2005)\citenamefont {Qiu},
  \citenamefont {Proffen}, \citenamefont {Mitchell},\ and\ \citenamefont
  {Billinge}}]{qiu;prl05}%
  \BibitemOpen
  \bibfield  {author} {\bibinfo {author} {\bibfnamefont {X.}~\bibnamefont
  {Qiu}}, \bibinfo {author} {\bibfnamefont {T.}~\bibnamefont {Proffen}},
  \bibinfo {author} {\bibfnamefont {J.~F.}\ \bibnamefont {Mitchell}}, \ and\
  \bibinfo {author} {\bibfnamefont {S.~J.~L.}\ \bibnamefont {Billinge}},\
  }\href@noop {} {\bibfield  {journal} {\bibinfo  {journal} {Phys. Rev. Lett.}\
  }\textbf {\bibinfo {volume} {94}},\ \bibinfo {pages} {177203} (\bibinfo
  {year} {2005})}\BibitemShut {NoStop}%
\bibitem [{\citenamefont {Bo\v{z}in}\ \emph {et~al.}(2008)\citenamefont
  {Bo\v{z}in}, \citenamefont {Sartbaeva}, \citenamefont {Zheng}, \citenamefont
  {Wells}, \citenamefont {Mitchell}, \citenamefont {Proffen}, \citenamefont
  {Thorpe},\ and\ \citenamefont {Billinge}}]{bozin;jpcs08}%
  \BibitemOpen
  \bibfield  {author} {\bibinfo {author} {\bibfnamefont {E.~S.}\ \bibnamefont
  {Bo\v{z}in}}, \bibinfo {author} {\bibfnamefont {A.}~\bibnamefont
  {Sartbaeva}}, \bibinfo {author} {\bibfnamefont {H.}~\bibnamefont {Zheng}},
  \bibinfo {author} {\bibfnamefont {S.~A.}\ \bibnamefont {Wells}}, \bibinfo
  {author} {\bibfnamefont {J.~F.}\ \bibnamefont {Mitchell}}, \bibinfo {author}
  {\bibfnamefont {T.}~\bibnamefont {Proffen}}, \bibinfo {author} {\bibfnamefont
  {M.~F.}\ \bibnamefont {Thorpe}}, \ and\ \bibinfo {author} {\bibfnamefont
  {S.~J.~L.}\ \bibnamefont {Billinge}},\ }\href {\doibase
  10.1016/j.jpcs.2008.03.029} {\bibfield  {journal} {\bibinfo  {journal} {J.
  Phys. Chem. Solids}\ }\textbf {\bibinfo {volume} {69}},\ \bibinfo {pages}
  {2146 } (\bibinfo {year} {2008})}\BibitemShut {NoStop}%
\bibitem [{\citenamefont {Billinge}\ \emph
  {et~al.}(1994{\natexlab{b}})\citenamefont {Billinge}, \citenamefont {Kwei},\
  and\ \citenamefont {Takagi}}]{billi;pb94}%
  \BibitemOpen
  \bibfield  {author} {\bibinfo {author} {\bibfnamefont {S.~J.~L.}\
  \bibnamefont {Billinge}}, \bibinfo {author} {\bibfnamefont {G.~H.}\
  \bibnamefont {Kwei}}, \ and\ \bibinfo {author} {\bibfnamefont
  {H.}~\bibnamefont {Takagi}},\ }\href@noop {} {\bibfield  {journal} {\bibinfo
  {journal} {Physica B}\ }\textbf {\bibinfo {volume} {199-200}},\ \bibinfo
  {pages} {244} (\bibinfo {year} {1994}{\natexlab{b}})}\BibitemShut {NoStop}%
\bibitem [{\citenamefont {Debye}(1912)}]{debye;adp12}%
  \BibitemOpen
  \bibfield  {author} {\bibinfo {author} {\bibfnamefont {P.}~\bibnamefont
  {Debye}},\ }\href@noop {} {\bibfield  {journal} {\bibinfo  {journal} {Ann.
  Phys.-Berlin}\ }\textbf {\bibinfo {volume} {39}},\ \bibinfo {pages} {789}
  (\bibinfo {year} {1912})}\BibitemShut {NoStop}%
\bibitem [{\citenamefont {Knox}\ \emph {et~al.}(2013)\citenamefont {Knox},
  \citenamefont {Abeykoon}, \citenamefont {Zheng}, \citenamefont {Yin},
  \citenamefont {Tsvelik}, \citenamefont {Mitchell}, \citenamefont {Billinge},\
  and\ \citenamefont {Bozin}}]{knox;prb13}%
  \BibitemOpen
  \bibfield  {author} {\bibinfo {author} {\bibfnamefont {K.~R.}\ \bibnamefont
  {Knox}}, \bibinfo {author} {\bibfnamefont {A.~M.~M.}\ \bibnamefont
  {Abeykoon}}, \bibinfo {author} {\bibfnamefont {H.}~\bibnamefont {Zheng}},
  \bibinfo {author} {\bibfnamefont {W.-G.}\ \bibnamefont {Yin}}, \bibinfo
  {author} {\bibfnamefont {A.~M.}\ \bibnamefont {Tsvelik}}, \bibinfo {author}
  {\bibfnamefont {J.~F.}\ \bibnamefont {Mitchell}}, \bibinfo {author}
  {\bibfnamefont {S.~J.~L.}\ \bibnamefont {Billinge}}, \ and\ \bibinfo {author}
  {\bibfnamefont {E.~S.}\ \bibnamefont {Bozin}},\ }\href {\doibase
  10.1103/PhysRevB.88.174114} {\bibfield  {journal} {\bibinfo  {journal} {Phys.
  Rev. B}\ }\textbf {\bibinfo {volume} {88}},\ \bibinfo {pages} {174114}
  (\bibinfo {year} {2013})}\BibitemShut {NoStop}%
\bibitem [{\citenamefont {Bo\v{z}in}\ \emph {et~al.}(2014)\citenamefont
  {Bo\v{z}in}, \citenamefont {Knox}, \citenamefont {Juh\'{a}s}, \citenamefont
  {Hor}, \citenamefont {Mitchell},\ and\ \citenamefont
  {Billinge}}]{bozin;sr14}%
  \BibitemOpen
  \bibfield  {author} {\bibinfo {author} {\bibfnamefont {E.~S.}\ \bibnamefont
  {Bo\v{z}in}}, \bibinfo {author} {\bibfnamefont {K.~R.}\ \bibnamefont {Knox}},
  \bibinfo {author} {\bibfnamefont {P.}~\bibnamefont {Juh\'{a}s}}, \bibinfo
  {author} {\bibfnamefont {Y.~S.}\ \bibnamefont {Hor}}, \bibinfo {author}
  {\bibfnamefont {J.~F.}\ \bibnamefont {Mitchell}}, \ and\ \bibinfo {author}
  {\bibfnamefont {S.~J.~L.}\ \bibnamefont {Billinge}},\ }\href {\doibase
  10.1038/srep04081} {\bibfield  {journal} {\bibinfo  {journal} {Sci. Rep.}\
  }\textbf {\bibinfo {volume} {4}},\ \bibinfo {pages} {4081} (\bibinfo {year}
  {2014})}\BibitemShut {NoStop}%
\bibitem [{\citenamefont {Billinge}\ \emph
  {et~al.}(1994{\natexlab{c}})\citenamefont {Billinge}, \citenamefont {Kwei},\
  and\ \citenamefont {Takagi}}]{billi;prl94}%
  \BibitemOpen
  \bibfield  {author} {\bibinfo {author} {\bibfnamefont {S.~J.~L.}\
  \bibnamefont {Billinge}}, \bibinfo {author} {\bibfnamefont {G.~H.}\
  \bibnamefont {Kwei}}, \ and\ \bibinfo {author} {\bibfnamefont
  {H.}~\bibnamefont {Takagi}},\ }\href@noop {} {\bibfield  {journal} {\bibinfo
  {journal} {Phys. Rev. Lett.}\ }\textbf {\bibinfo {volume} {72}},\ \bibinfo
  {pages} {2282} (\bibinfo {year} {1994}{\natexlab{c}})}\BibitemShut {NoStop}%
\bibitem [{\citenamefont {Bo{\v z}in}\ and\ \citenamefont
  {Billinge}(1998)}]{bozin;ssp98}%
  \BibitemOpen
  \bibfield  {author} {\bibinfo {author} {\bibfnamefont {E.~S.}\ \bibnamefont
  {Bo{\v z}in}}\ and\ \bibinfo {author} {\bibfnamefont {S.~J.~L.}\ \bibnamefont
  {Billinge}},\ }\href@noop {} {\bibfield  {journal} {\bibinfo  {journal}
  {Solid State Phenomena}\ }\textbf {\bibinfo {volume} {61-62}},\ \bibinfo
  {pages} {271} (\bibinfo {year} {1998})}\BibitemShut {NoStop}%
\bibitem [{\citenamefont {{Bo\v zin}}\ \emph {et~al.}(1998)\citenamefont {{Bo\v
  zin}}, \citenamefont {Billinge},\ and\ \citenamefont {Kwei}}]{bozin;pb98}%
  \BibitemOpen
  \bibfield  {author} {\bibinfo {author} {\bibfnamefont {E.~S.}\ \bibnamefont
  {{Bo\v zin}}}, \bibinfo {author} {\bibfnamefont {S.~J.~L.}\ \bibnamefont
  {Billinge}}, \ and\ \bibinfo {author} {\bibfnamefont {G.~H.}\ \bibnamefont
  {Kwei}},\ }\href@noop {} {\bibfield  {journal} {\bibinfo  {journal} {Physica
  B}\ }\textbf {\bibinfo {volume} {241-243}},\ \bibinfo {pages} {795} (\bibinfo
  {year} {1998})}\BibitemShut {NoStop}%
\bibitem [{\citenamefont {Bo{\v z}in}\ \emph {et~al.}(1999)\citenamefont {Bo{\v
  z}in}, \citenamefont {Billinge}, \citenamefont {Kwei},\ and\ \citenamefont
  {Takagi}}]{bozin;prb99}%
  \BibitemOpen
  \bibfield  {author} {\bibinfo {author} {\bibfnamefont {E.~S.}\ \bibnamefont
  {Bo{\v z}in}}, \bibinfo {author} {\bibfnamefont {S.~J.~L.}\ \bibnamefont
  {Billinge}}, \bibinfo {author} {\bibfnamefont {G.~H.}\ \bibnamefont {Kwei}},
  \ and\ \bibinfo {author} {\bibfnamefont {H.}~\bibnamefont {Takagi}},\
  }\href@noop {} {\bibfield  {journal} {\bibinfo  {journal} {Phys. Rev. B}\
  }\textbf {\bibinfo {volume} {59}},\ \bibinfo {pages} {4445} (\bibinfo {year}
  {1999})}\BibitemShut {NoStop}%
\bibitem [{\citenamefont {{Bo\v zin}}\ \emph {et~al.}(2010)\citenamefont {{Bo\v
  zin}}, \citenamefont {Malliakas}, \citenamefont {Souvatzis}, \citenamefont
  {Proffen}, \citenamefont {Spaldin}, \citenamefont {Kanatzidis},\ and\
  \citenamefont {Billinge}}]{bozin;s10}%
  \BibitemOpen
  \bibfield  {author} {\bibinfo {author} {\bibfnamefont {E.~S.}\ \bibnamefont
  {{Bo\v zin}}}, \bibinfo {author} {\bibfnamefont {C.~D.}\ \bibnamefont
  {Malliakas}}, \bibinfo {author} {\bibfnamefont {P.}~\bibnamefont
  {Souvatzis}}, \bibinfo {author} {\bibfnamefont {T.}~\bibnamefont {Proffen}},
  \bibinfo {author} {\bibfnamefont {N.~A.}\ \bibnamefont {Spaldin}}, \bibinfo
  {author} {\bibfnamefont {M.~G.}\ \bibnamefont {Kanatzidis}}, \ and\ \bibinfo
  {author} {\bibfnamefont {S.~J.~L.}\ \bibnamefont {Billinge}},\ }\href@noop {}
  {\bibfield  {journal} {\bibinfo  {journal} {Science}\ }\textbf {\bibinfo
  {volume} {330}},\ \bibinfo {pages} {1660} (\bibinfo {year}
  {2010})}\BibitemShut {NoStop}%
\bibitem [{\citenamefont {Knox}\ \emph {et~al.}(2014)\citenamefont {Knox},
  \citenamefont {Bozin}, \citenamefont {Malliakas}, \citenamefont
  {Kanatzidis},\ and\ \citenamefont {Billinge}}]{knox;prb14}%
  \BibitemOpen
  \bibfield  {author} {\bibinfo {author} {\bibfnamefont {K.~R.}\ \bibnamefont
  {Knox}}, \bibinfo {author} {\bibfnamefont {E.~S.}\ \bibnamefont {Bozin}},
  \bibinfo {author} {\bibfnamefont {C.~D.}\ \bibnamefont {Malliakas}}, \bibinfo
  {author} {\bibfnamefont {M.~G.}\ \bibnamefont {Kanatzidis}}, \ and\ \bibinfo
  {author} {\bibfnamefont {S.~J.~L.}\ \bibnamefont {Billinge}},\ }\href
  {\doibase 10.1103/PhysRevB.89.014102} {\bibfield  {journal} {\bibinfo
  {journal} {Phys. Rev. B}\ }\textbf {\bibinfo {volume} {89}},\ \bibinfo
  {pages} {014102} (\bibinfo {year} {2014})}\BibitemShut {NoStop}%
\bibitem [{\citenamefont {Jeong}\ \emph {et~al.}(1999)\citenamefont {Jeong},
  \citenamefont {Proffen}, \citenamefont {Mohiuddin-Jacobs},\ and\
  \citenamefont {Billinge}}]{jeong;jpc99}%
  \BibitemOpen
  \bibfield  {author} {\bibinfo {author} {\bibfnamefont {I.}~\bibnamefont
  {Jeong}}, \bibinfo {author} {\bibfnamefont {T.}~\bibnamefont {Proffen}},
  \bibinfo {author} {\bibfnamefont {F.}~\bibnamefont {Mohiuddin-Jacobs}}, \
  and\ \bibinfo {author} {\bibfnamefont {S.~J.~L.}\ \bibnamefont {Billinge}},\
  }\href@noop {} {\bibfield  {journal} {\bibinfo  {journal} {J. Phys. Chem. A}\
  }\textbf {\bibinfo {volume} {103}},\ \bibinfo {pages} {921} (\bibinfo {year}
  {1999})}\BibitemShut {NoStop}%
\bibitem [{\citenamefont {Proffen}\ and\ \citenamefont
  {Billinge}(1999)}]{proff;jac99}%
  \BibitemOpen
  \bibfield  {author} {\bibinfo {author} {\bibfnamefont {T.}~\bibnamefont
  {Proffen}}\ and\ \bibinfo {author} {\bibfnamefont {S.~J.~L.}\ \bibnamefont
  {Billinge}},\ }\href@noop {} {\bibfield  {journal} {\bibinfo  {journal} {J.
  Appl. Crystallogr.}\ }\textbf {\bibinfo {volume} {32}},\ \bibinfo {pages}
  {572} (\bibinfo {year} {1999})}\BibitemShut {NoStop}%
\bibitem [{\citenamefont {Isaacs}\ \emph {et~al.}(1994)\citenamefont {Isaacs},
  \citenamefont {Aeppli}, \citenamefont {Zschack}, \citenamefont {Cheong},
  \citenamefont {Williams},\ and\ \citenamefont {Buttrey}}]{isaac;prl94}%
  \BibitemOpen
  \bibfield  {author} {\bibinfo {author} {\bibfnamefont {E.~D.}\ \bibnamefont
  {Isaacs}}, \bibinfo {author} {\bibfnamefont {G.}~\bibnamefont {Aeppli}},
  \bibinfo {author} {\bibfnamefont {P.}~\bibnamefont {Zschack}}, \bibinfo
  {author} {\bibfnamefont {S.-W.}\ \bibnamefont {Cheong}}, \bibinfo {author}
  {\bibfnamefont {H.}~\bibnamefont {Williams}}, \ and\ \bibinfo {author}
  {\bibfnamefont {D.~J.}\ \bibnamefont {Buttrey}},\ }\href@noop {} {\bibfield
  {journal} {\bibinfo  {journal} {Phys. Rev. Lett.}\ }\textbf {\bibinfo
  {volume} {72}},\ \bibinfo {pages} {3421} (\bibinfo {year}
  {1994})}\BibitemShut {NoStop}%
\bibitem [{\citenamefont {Bianconi}\ \emph {et~al.}(1996)\citenamefont
  {Bianconi}, \citenamefont {Saini}, \citenamefont {Lanzara}, \citenamefont
  {Missori}, \citenamefont {Oyanagi}, \citenamefont {Yamaguchi}, \citenamefont
  {Oka},\ and\ \citenamefont {Ito}}]{bian96}%
  \BibitemOpen
  \bibfield  {author} {\bibinfo {author} {\bibfnamefont {A.}~\bibnamefont
  {Bianconi}}, \bibinfo {author} {\bibfnamefont {N.~L.}\ \bibnamefont {Saini}},
  \bibinfo {author} {\bibfnamefont {A.}~\bibnamefont {Lanzara}}, \bibinfo
  {author} {\bibfnamefont {M.}~\bibnamefont {Missori}}, \bibinfo {author}
  {\bibfnamefont {T.~R.~H.}\ \bibnamefont {Oyanagi}}, \bibinfo {author}
  {\bibfnamefont {H.}~\bibnamefont {Yamaguchi}}, \bibinfo {author}
  {\bibfnamefont {K.}~\bibnamefont {Oka}}, \ and\ \bibinfo {author}
  {\bibfnamefont {T.}~\bibnamefont {Ito}},\ }\href@noop {} {\bibfield
  {journal} {\bibinfo  {journal} {Phys. Rev. Lett.}\ }\textbf {\bibinfo
  {volume} {76}},\ \bibinfo {pages} {3412} (\bibinfo {year}
  {1996})}\BibitemShut {NoStop}%
\bibitem [{\citenamefont {Saini}\ \emph {et~al.}(1997)\citenamefont {Saini},
  \citenamefont {Lanzara}, \citenamefont {Oyanagi}, \citenamefont {Yamaguchi},
  \citenamefont {Oka}, \citenamefont {Ito},\ and\ \citenamefont
  {Bianconi}}]{sain97}%
  \BibitemOpen
  \bibfield  {author} {\bibinfo {author} {\bibfnamefont {N.~L.}\ \bibnamefont
  {Saini}}, \bibinfo {author} {\bibfnamefont {A.}~\bibnamefont {Lanzara}},
  \bibinfo {author} {\bibfnamefont {H.}~\bibnamefont {Oyanagi}}, \bibinfo
  {author} {\bibfnamefont {H.}~\bibnamefont {Yamaguchi}}, \bibinfo {author}
  {\bibfnamefont {K.}~\bibnamefont {Oka}}, \bibinfo {author} {\bibfnamefont
  {T.}~\bibnamefont {Ito}}, \ and\ \bibinfo {author} {\bibfnamefont
  {A.}~\bibnamefont {Bianconi}},\ }\href@noop {} {\bibfield  {journal}
  {\bibinfo  {journal} {Phys. Rev. B}\ }\textbf {\bibinfo {volume} {55}},\
  \bibinfo {pages} {12759} (\bibinfo {year} {1997})}\BibitemShut {NoStop}%
\bibitem [{\citenamefont {Lanzara}\ \emph {et~al.}(1996)\citenamefont
  {Lanzara}, \citenamefont {Saini}, \citenamefont {Rossetti}, \citenamefont
  {Bianconi}, \citenamefont {Oyanagi}, \citenamefont {Yamaguchi},\ and\
  \citenamefont {Maeno}}]{lanz96}%
  \BibitemOpen
  \bibfield  {author} {\bibinfo {author} {\bibfnamefont {A.}~\bibnamefont
  {Lanzara}}, \bibinfo {author} {\bibfnamefont {N.~L.}\ \bibnamefont {Saini}},
  \bibinfo {author} {\bibfnamefont {T.}~\bibnamefont {Rossetti}}, \bibinfo
  {author} {\bibfnamefont {A.}~\bibnamefont {Bianconi}}, \bibinfo {author}
  {\bibfnamefont {H.}~\bibnamefont {Oyanagi}}, \bibinfo {author} {\bibfnamefont
  {H.}~\bibnamefont {Yamaguchi}}, \ and\ \bibinfo {author} {\bibfnamefont
  {Y.}~\bibnamefont {Maeno}},\ }\href@noop {} {\bibfield  {journal} {\bibinfo
  {journal} {Solid State Commun.}\ }\textbf {\bibinfo {volume} {97}},\ \bibinfo
  {pages} {93} (\bibinfo {year} {1996})}\BibitemShut {NoStop}%
\bibitem [{\citenamefont {Tranquada}\ \emph {et~al.}(1987)\citenamefont
  {Tranquada}, \citenamefont {Heald},\ and\ \citenamefont
  {Moodenbaugh}}]{tran87b}%
  \BibitemOpen
  \bibfield  {author} {\bibinfo {author} {\bibfnamefont {J.~M.}\ \bibnamefont
  {Tranquada}}, \bibinfo {author} {\bibfnamefont {S.~M.}\ \bibnamefont
  {Heald}}, \ and\ \bibinfo {author} {\bibfnamefont {A.~R.}\ \bibnamefont
  {Moodenbaugh}},\ }\href@noop {} {\bibfield  {journal} {\bibinfo  {journal}
  {Phys. Rev. B}\ }\textbf {\bibinfo {volume} {36}},\ \bibinfo {pages} {8401}
  (\bibinfo {year} {1987})}\BibitemShut {NoStop}%
\bibitem [{\citenamefont {Tranquada}\ \emph {et~al.}(1996)\citenamefont
  {Tranquada}, \citenamefont {Axe}, \citenamefont {Ichikawa}, \citenamefont
  {Nakamura}, \citenamefont {Uchida},\ and\ \citenamefont {Nachumi}}]{tran96b}%
  \BibitemOpen
  \bibfield  {author} {\bibinfo {author} {\bibfnamefont {J.~M.}\ \bibnamefont
  {Tranquada}}, \bibinfo {author} {\bibfnamefont {J.~D.}\ \bibnamefont {Axe}},
  \bibinfo {author} {\bibfnamefont {N.}~\bibnamefont {Ichikawa}}, \bibinfo
  {author} {\bibfnamefont {Y.}~\bibnamefont {Nakamura}}, \bibinfo {author}
  {\bibfnamefont {S.}~\bibnamefont {Uchida}}, \ and\ \bibinfo {author}
  {\bibfnamefont {B.}~\bibnamefont {Nachumi}},\ }\href@noop {} {\bibfield
  {journal} {\bibinfo  {journal} {Phys. Rev. B}\ }\textbf {\bibinfo {volume}
  {54}},\ \bibinfo {pages} {7489} (\bibinfo {year} {1996})}\BibitemShut
  {NoStop}%
\bibitem [{\citenamefont {Suzuki}\ \emph {et~al.}(1994)\citenamefont {Suzuki},
  \citenamefont {Sera}, \citenamefont {Hanaguri},\ and\ \citenamefont
  {Fukase}}]{suzu94}%
  \BibitemOpen
  \bibfield  {author} {\bibinfo {author} {\bibfnamefont {T.}~\bibnamefont
  {Suzuki}}, \bibinfo {author} {\bibfnamefont {M.}~\bibnamefont {Sera}},
  \bibinfo {author} {\bibfnamefont {T.}~\bibnamefont {Hanaguri}}, \ and\
  \bibinfo {author} {\bibfnamefont {T.}~\bibnamefont {Fukase}},\ }\href@noop {}
  {\bibfield  {journal} {\bibinfo  {journal} {Phys Rev. B}\ }\textbf {\bibinfo
  {volume} {49}},\ \bibinfo {pages} {12392} (\bibinfo {year}
  {1994})}\BibitemShut {NoStop}%
\bibitem [{\citenamefont {Fausti}\ \emph {et~al.}(2011)\citenamefont {Fausti},
  \citenamefont {Tobey}, \citenamefont {Dean}, \citenamefont {Kaiser},
  \citenamefont {Dienst}, \citenamefont {Hoffmann}, \citenamefont {Pyon},
  \citenamefont {Takayama}, \citenamefont {Takagi},\ and\ \citenamefont
  {Cavalleri}}]{faus11}%
  \BibitemOpen
  \bibfield  {author} {\bibinfo {author} {\bibfnamefont {D.}~\bibnamefont
  {Fausti}}, \bibinfo {author} {\bibfnamefont {R.~I.}\ \bibnamefont {Tobey}},
  \bibinfo {author} {\bibfnamefont {N.}~\bibnamefont {Dean}}, \bibinfo {author}
  {\bibfnamefont {S.}~\bibnamefont {Kaiser}}, \bibinfo {author} {\bibfnamefont
  {A.}~\bibnamefont {Dienst}}, \bibinfo {author} {\bibfnamefont {M.~C.}\
  \bibnamefont {Hoffmann}}, \bibinfo {author} {\bibfnamefont {S.}~\bibnamefont
  {Pyon}}, \bibinfo {author} {\bibfnamefont {T.}~\bibnamefont {Takayama}},
  \bibinfo {author} {\bibfnamefont {H.}~\bibnamefont {Takagi}}, \ and\ \bibinfo
  {author} {\bibfnamefont {A.}~\bibnamefont {Cavalleri}},\ }\href@noop {}
  {\bibfield  {journal} {\bibinfo  {journal} {Science}\ }\textbf {\bibinfo
  {volume} {331}},\ \bibinfo {pages} {189} (\bibinfo {year}
  {2011})}\BibitemShut {NoStop}%
\bibitem [{\citenamefont {{F\"orst}}\ \emph {et~al.}(2014)\citenamefont
  {{F\"orst}}, \citenamefont {Tobey}, \citenamefont {Bromberger}, \citenamefont
  {Wilkins}, \citenamefont {Khanna}, \citenamefont {Caviglia}, \citenamefont
  {Chuang}, \citenamefont {Lee}, \citenamefont {Schlotter}, \citenamefont
  {Turner}, \citenamefont {Minitti}, \citenamefont {Krupin}, \citenamefont
  {Xu}, \citenamefont {Wen}, \citenamefont {Gu}, \citenamefont {Dhesi},
  \citenamefont {Cavalleri},\ and\ \citenamefont {Hill}}]{fors14}%
  \BibitemOpen
  \bibfield  {author} {\bibinfo {author} {\bibfnamefont {M.}~\bibnamefont
  {{F\"orst}}}, \bibinfo {author} {\bibfnamefont {R.~I.}\ \bibnamefont
  {Tobey}}, \bibinfo {author} {\bibfnamefont {H.}~\bibnamefont {Bromberger}},
  \bibinfo {author} {\bibfnamefont {S.~B.}\ \bibnamefont {Wilkins}}, \bibinfo
  {author} {\bibfnamefont {V.}~\bibnamefont {Khanna}}, \bibinfo {author}
  {\bibfnamefont {A.~D.}\ \bibnamefont {Caviglia}}, \bibinfo {author}
  {\bibfnamefont {Y.-D.}\ \bibnamefont {Chuang}}, \bibinfo {author}
  {\bibfnamefont {W.~S.}\ \bibnamefont {Lee}}, \bibinfo {author} {\bibfnamefont
  {W.~F.}\ \bibnamefont {Schlotter}}, \bibinfo {author} {\bibfnamefont {J.~J.}\
  \bibnamefont {Turner}}, \bibinfo {author} {\bibfnamefont {M.~P.}\
  \bibnamefont {Minitti}}, \bibinfo {author} {\bibfnamefont {O.}~\bibnamefont
  {Krupin}}, \bibinfo {author} {\bibfnamefont {Z.~J.}\ \bibnamefont {Xu}},
  \bibinfo {author} {\bibfnamefont {J.~S.}\ \bibnamefont {Wen}}, \bibinfo
  {author} {\bibfnamefont {G.~D.}\ \bibnamefont {Gu}}, \bibinfo {author}
  {\bibfnamefont {S.~S.}\ \bibnamefont {Dhesi}}, \bibinfo {author}
  {\bibfnamefont {A.}~\bibnamefont {Cavalleri}}, \ and\ \bibinfo {author}
  {\bibfnamefont {J.~P.}\ \bibnamefont {Hill}},\ }\href@noop {} {\bibfield
  {journal} {\bibinfo  {journal} {Phys. Rev. Lett.}\ }\textbf {\bibinfo
  {volume} {112}},\ \bibinfo {pages} {157002} (\bibinfo {year}
  {2014})}\BibitemShut {NoStop}%
\bibitem [{\citenamefont {Nicoletti}\ \emph {et~al.}(2014)\citenamefont
  {Nicoletti}, \citenamefont {Casandruc}, \citenamefont {Laplace},
  \citenamefont {Khanna}, \citenamefont {Hunt}, \citenamefont {Kaiser},
  \citenamefont {Dhesi}, \citenamefont {Gu}, \citenamefont {Hill},\ and\
  \citenamefont {Cavalleri}}]{nico14}%
  \BibitemOpen
  \bibfield  {author} {\bibinfo {author} {\bibfnamefont {D.}~\bibnamefont
  {Nicoletti}}, \bibinfo {author} {\bibfnamefont {E.}~\bibnamefont
  {Casandruc}}, \bibinfo {author} {\bibfnamefont {Y.}~\bibnamefont {Laplace}},
  \bibinfo {author} {\bibfnamefont {V.}~\bibnamefont {Khanna}}, \bibinfo
  {author} {\bibfnamefont {C.~R.}\ \bibnamefont {Hunt}}, \bibinfo {author}
  {\bibfnamefont {S.}~\bibnamefont {Kaiser}}, \bibinfo {author} {\bibfnamefont
  {S.~S.}\ \bibnamefont {Dhesi}}, \bibinfo {author} {\bibfnamefont {G.~D.}\
  \bibnamefont {Gu}}, \bibinfo {author} {\bibfnamefont {J.~P.}\ \bibnamefont
  {Hill}}, \ and\ \bibinfo {author} {\bibfnamefont {A.}~\bibnamefont
  {Cavalleri}},\ }\href {\doibase 10.1103/PhysRevB.90.100503} {\bibfield
  {journal} {\bibinfo  {journal} {Phys. Rev. B}\ }\textbf {\bibinfo {volume}
  {90}},\ \bibinfo {pages} {100503} (\bibinfo {year} {2014})}\BibitemShut
  {NoStop}%
\bibitem [{\citenamefont {Ghiringhelli}\ \emph {et~al.}(2012)\citenamefont
  {Ghiringhelli}, \citenamefont {Le~Tacon}, \citenamefont {Minola},
  \citenamefont {Blanco-Canosa}, \citenamefont {Mazzoli}, \citenamefont
  {Brookes}, \citenamefont {De~Luca}, \citenamefont {Frano}, \citenamefont
  {Hawthorn}, \citenamefont {He}, \citenamefont {Loew}, \citenamefont {Sala},
  \citenamefont {Peets}, \citenamefont {Salluzzo}, \citenamefont {Schierle},
  \citenamefont {Sutarto}, \citenamefont {Sawatzky}, \citenamefont {Weschke},
  \citenamefont {Keimer},\ and\ \citenamefont {Braicovich}}]{ghir12}%
  \BibitemOpen
  \bibfield  {author} {\bibinfo {author} {\bibfnamefont {G.}~\bibnamefont
  {Ghiringhelli}}, \bibinfo {author} {\bibfnamefont {M.}~\bibnamefont
  {Le~Tacon}}, \bibinfo {author} {\bibfnamefont {M.}~\bibnamefont {Minola}},
  \bibinfo {author} {\bibfnamefont {S.}~\bibnamefont {Blanco-Canosa}}, \bibinfo
  {author} {\bibfnamefont {C.}~\bibnamefont {Mazzoli}}, \bibinfo {author}
  {\bibfnamefont {N.~B.}\ \bibnamefont {Brookes}}, \bibinfo {author}
  {\bibfnamefont {G.~M.}\ \bibnamefont {De~Luca}}, \bibinfo {author}
  {\bibfnamefont {A.}~\bibnamefont {Frano}}, \bibinfo {author} {\bibfnamefont
  {D.~G.}\ \bibnamefont {Hawthorn}}, \bibinfo {author} {\bibfnamefont
  {F.}~\bibnamefont {He}}, \bibinfo {author} {\bibfnamefont {T.}~\bibnamefont
  {Loew}}, \bibinfo {author} {\bibfnamefont {M.~M.}\ \bibnamefont {Sala}},
  \bibinfo {author} {\bibfnamefont {D.~C.}\ \bibnamefont {Peets}}, \bibinfo
  {author} {\bibfnamefont {M.}~\bibnamefont {Salluzzo}}, \bibinfo {author}
  {\bibfnamefont {E.}~\bibnamefont {Schierle}}, \bibinfo {author}
  {\bibfnamefont {R.}~\bibnamefont {Sutarto}}, \bibinfo {author} {\bibfnamefont
  {G.~A.}\ \bibnamefont {Sawatzky}}, \bibinfo {author} {\bibfnamefont
  {E.}~\bibnamefont {Weschke}}, \bibinfo {author} {\bibfnamefont
  {B.}~\bibnamefont {Keimer}}, \ and\ \bibinfo {author} {\bibfnamefont
  {L.}~\bibnamefont {Braicovich}},\ }\href@noop {} {\bibfield  {journal}
  {\bibinfo  {journal} {Science}\ }\textbf {\bibinfo {volume} {337}},\ \bibinfo
  {pages} {821} (\bibinfo {year} {2012})}\BibitemShut {NoStop}%
\bibitem [{\citenamefont {Chang}\ \emph {et~al.}(2012)\citenamefont {Chang},
  \citenamefont {Blackburn}, \citenamefont {Holmes}, \citenamefont
  {Christensen}, \citenamefont {Larsen}, \citenamefont {Mesot}, \citenamefont
  {Liang}, \citenamefont {Bonn}, \citenamefont {Hardy}, \citenamefont
  {Watenphul}, \citenamefont {Zimmermann}, \citenamefont {Forgan},\ and\
  \citenamefont {Hayden}}]{chan12a}%
  \BibitemOpen
  \bibfield  {author} {\bibinfo {author} {\bibfnamefont {J.}~\bibnamefont
  {Chang}}, \bibinfo {author} {\bibfnamefont {E.}~\bibnamefont {Blackburn}},
  \bibinfo {author} {\bibfnamefont {A.~T.}\ \bibnamefont {Holmes}}, \bibinfo
  {author} {\bibfnamefont {N.~B.}\ \bibnamefont {Christensen}}, \bibinfo
  {author} {\bibfnamefont {J.}~\bibnamefont {Larsen}}, \bibinfo {author}
  {\bibfnamefont {J.}~\bibnamefont {Mesot}}, \bibinfo {author} {\bibfnamefont
  {R.}~\bibnamefont {Liang}}, \bibinfo {author} {\bibfnamefont {D.~A.}\
  \bibnamefont {Bonn}}, \bibinfo {author} {\bibfnamefont {W.~N.}\ \bibnamefont
  {Hardy}}, \bibinfo {author} {\bibfnamefont {A.}~\bibnamefont {Watenphul}},
  \bibinfo {author} {\bibfnamefont {M.~v.}\ \bibnamefont {Zimmermann}},
  \bibinfo {author} {\bibfnamefont {E.~M.}\ \bibnamefont {Forgan}}, \ and\
  \bibinfo {author} {\bibfnamefont {S.~M.}\ \bibnamefont {Hayden}},\
  }\href@noop {} {\bibfield  {journal} {\bibinfo  {journal} {Nat. Phys.}\
  }\textbf {\bibinfo {volume} {8}},\ \bibinfo {pages} {871} (\bibinfo {year}
  {2012})}\BibitemShut {NoStop}%
\bibitem [{\citenamefont {H\"ucker}\ \emph {et~al.}(2014)\citenamefont
  {H\"ucker}, \citenamefont {Christensen}, \citenamefont {Holmes},
  \citenamefont {Blackburn}, \citenamefont {Forgan}, \citenamefont {Liang},
  \citenamefont {Bonn}, \citenamefont {Hardy}, \citenamefont {Gutowski},
  \citenamefont {Zimmermann}, \citenamefont {Hayden},\ and\ \citenamefont
  {Chang}}]{huck14}%
  \BibitemOpen
  \bibfield  {author} {\bibinfo {author} {\bibfnamefont {M.}~\bibnamefont
  {H\"ucker}}, \bibinfo {author} {\bibfnamefont {N.~B.}\ \bibnamefont
  {Christensen}}, \bibinfo {author} {\bibfnamefont {A.~T.}\ \bibnamefont
  {Holmes}}, \bibinfo {author} {\bibfnamefont {E.}~\bibnamefont {Blackburn}},
  \bibinfo {author} {\bibfnamefont {E.~M.}\ \bibnamefont {Forgan}}, \bibinfo
  {author} {\bibfnamefont {R.}~\bibnamefont {Liang}}, \bibinfo {author}
  {\bibfnamefont {D.~A.}\ \bibnamefont {Bonn}}, \bibinfo {author}
  {\bibfnamefont {W.~N.}\ \bibnamefont {Hardy}}, \bibinfo {author}
  {\bibfnamefont {O.}~\bibnamefont {Gutowski}}, \bibinfo {author}
  {\bibfnamefont {M.~v.}\ \bibnamefont {Zimmermann}}, \bibinfo {author}
  {\bibfnamefont {S.~M.}\ \bibnamefont {Hayden}}, \ and\ \bibinfo {author}
  {\bibfnamefont {J.}~\bibnamefont {Chang}},\ }\href@noop {} {\bibfield
  {journal} {\bibinfo  {journal} {Phys. Rev. B}\ }\textbf {\bibinfo {volume}
  {90}},\ \bibinfo {pages} {054514} (\bibinfo {year} {2014})}\BibitemShut
  {NoStop}%
\bibitem [{\citenamefont {Blanco-Canosa}\ \emph {et~al.}(2014)\citenamefont
  {Blanco-Canosa}, \citenamefont {Frano}, \citenamefont {Schierle},
  \citenamefont {Porras}, \citenamefont {Loew}, \citenamefont {Minola},
  \citenamefont {Bluschke}, \citenamefont {Weschke}, \citenamefont {Keimer},\
  and\ \citenamefont {Le~Tacon}}]{blan14}%
  \BibitemOpen
  \bibfield  {author} {\bibinfo {author} {\bibfnamefont {S.}~\bibnamefont
  {Blanco-Canosa}}, \bibinfo {author} {\bibfnamefont {A.}~\bibnamefont
  {Frano}}, \bibinfo {author} {\bibfnamefont {E.}~\bibnamefont {Schierle}},
  \bibinfo {author} {\bibfnamefont {J.}~\bibnamefont {Porras}}, \bibinfo
  {author} {\bibfnamefont {T.}~\bibnamefont {Loew}}, \bibinfo {author}
  {\bibfnamefont {M.}~\bibnamefont {Minola}}, \bibinfo {author} {\bibfnamefont
  {M.}~\bibnamefont {Bluschke}}, \bibinfo {author} {\bibfnamefont
  {E.}~\bibnamefont {Weschke}}, \bibinfo {author} {\bibfnamefont
  {B.}~\bibnamefont {Keimer}}, \ and\ \bibinfo {author} {\bibfnamefont
  {M.}~\bibnamefont {Le~Tacon}},\ }\href@noop {} {\bibfield  {journal}
  {\bibinfo  {journal} {Phys. Rev. B}\ }\textbf {\bibinfo {volume} {90}},\
  \bibinfo {pages} {054513} (\bibinfo {year} {2014})}\BibitemShut {NoStop}%
\bibitem [{\citenamefont {Ichikawa}\ \emph {et~al.}(2000)\citenamefont
  {Ichikawa}, \citenamefont {Uchida}, \citenamefont {Tranquada}, \citenamefont
  {Niem\"oller}, \citenamefont {Gehring}, \citenamefont {Lee},\ and\
  \citenamefont {Schneider}}]{ichi00}%
  \BibitemOpen
  \bibfield  {author} {\bibinfo {author} {\bibfnamefont {N.}~\bibnamefont
  {Ichikawa}}, \bibinfo {author} {\bibfnamefont {S.}~\bibnamefont {Uchida}},
  \bibinfo {author} {\bibfnamefont {J.~M.}\ \bibnamefont {Tranquada}}, \bibinfo
  {author} {\bibfnamefont {T.}~\bibnamefont {Niem\"oller}}, \bibinfo {author}
  {\bibfnamefont {P.~M.}\ \bibnamefont {Gehring}}, \bibinfo {author}
  {\bibfnamefont {S.-H.}\ \bibnamefont {Lee}}, \ and\ \bibinfo {author}
  {\bibfnamefont {J.~R.}\ \bibnamefont {Schneider}},\ }\href@noop {} {\bibfield
   {journal} {\bibinfo  {journal} {Phys. Rev. Lett.}\ }\textbf {\bibinfo
  {volume} {85}},\ \bibinfo {pages} {1738} (\bibinfo {year}
  {2000})}\BibitemShut {NoStop}%
\bibitem [{\citenamefont {Wu}\ \emph {et~al.}(2012)\citenamefont {Wu},
  \citenamefont {Buchholz}, \citenamefont {Trabant}, \citenamefont {Chang},
  \citenamefont {Komarek}, \citenamefont {Heigl}, \citenamefont {Zimmermann},
  \citenamefont {Cwik}, \citenamefont {Nakamura}, \citenamefont {Braden},\ and\
  \citenamefont {Sch\"u{\ss}ler-Langeheine}}]{wu12}%
  \BibitemOpen
  \bibfield  {author} {\bibinfo {author} {\bibfnamefont {H.~H.}\ \bibnamefont
  {Wu}}, \bibinfo {author} {\bibfnamefont {M.}~\bibnamefont {Buchholz}},
  \bibinfo {author} {\bibfnamefont {C.}~\bibnamefont {Trabant}}, \bibinfo
  {author} {\bibfnamefont {C.~F.}\ \bibnamefont {Chang}}, \bibinfo {author}
  {\bibfnamefont {A.~C.}\ \bibnamefont {Komarek}}, \bibinfo {author}
  {\bibfnamefont {F.}~\bibnamefont {Heigl}}, \bibinfo {author} {\bibfnamefont
  {M.~v.}\ \bibnamefont {Zimmermann}}, \bibinfo {author} {\bibfnamefont
  {M.}~\bibnamefont {Cwik}}, \bibinfo {author} {\bibfnamefont {F.}~\bibnamefont
  {Nakamura}}, \bibinfo {author} {\bibfnamefont {M.}~\bibnamefont {Braden}}, \
  and\ \bibinfo {author} {\bibfnamefont {C.}~\bibnamefont
  {Sch\"u{\ss}ler-Langeheine}},\ }\href@noop {} {\bibfield  {journal} {\bibinfo
   {journal} {Nat. Commun.}\ }\textbf {\bibinfo {volume} {3}},\ \bibinfo
  {pages} {1023} (\bibinfo {year} {2012})}\BibitemShut {NoStop}%
\bibitem [{\citenamefont {Christensen}\ \emph {et~al.}(2014)\citenamefont
  {Christensen}, \citenamefont {Chang}, \citenamefont {Larsen}, \citenamefont
  {Fujita}, \citenamefont {Oda}, \citenamefont {Ido}, \citenamefont {Momono},
  \citenamefont {Forgan}, \citenamefont {Holmes}, \citenamefont {Mesot},
  \citenamefont {H\"ucker},\ and\ \citenamefont {v.~Zimmermann}}]{chri14}%
  \BibitemOpen
  \bibfield  {author} {\bibinfo {author} {\bibfnamefont {N.~B.}\ \bibnamefont
  {Christensen}}, \bibinfo {author} {\bibfnamefont {J.}~\bibnamefont {Chang}},
  \bibinfo {author} {\bibfnamefont {J.}~\bibnamefont {Larsen}}, \bibinfo
  {author} {\bibfnamefont {M.}~\bibnamefont {Fujita}}, \bibinfo {author}
  {\bibfnamefont {M.}~\bibnamefont {Oda}}, \bibinfo {author} {\bibfnamefont
  {M.}~\bibnamefont {Ido}}, \bibinfo {author} {\bibfnamefont {N.}~\bibnamefont
  {Momono}}, \bibinfo {author} {\bibfnamefont {E.~M.}\ \bibnamefont {Forgan}},
  \bibinfo {author} {\bibfnamefont {A.~T.}\ \bibnamefont {Holmes}}, \bibinfo
  {author} {\bibfnamefont {J.}~\bibnamefont {Mesot}}, \bibinfo {author}
  {\bibfnamefont {M.}~\bibnamefont {H\"ucker}}, \ and\ \bibinfo {author}
  {\bibfnamefont {M.}~\bibnamefont {v.~Zimmermann}},\ }\href@noop {} {\enquote
  {\bibinfo {title} {{Bulk charge stripe order competing with superconductivity
  in La$_{2-x}$Sr$_x$CuO$_4$ ($x=0.12$)}},}\ }\bibinfo {howpublished}
  {arXiv:1404.3192} (\bibinfo {year} {2014})\BibitemShut {NoStop}%
\bibitem [{\citenamefont {Croft}\ \emph {et~al.}(2014)\citenamefont {Croft},
  \citenamefont {Lester}, \citenamefont {Senn}, \citenamefont {Bombardi},\ and\
  \citenamefont {Hayden}}]{crof14}%
  \BibitemOpen
  \bibfield  {author} {\bibinfo {author} {\bibfnamefont {T.~P.}\ \bibnamefont
  {Croft}}, \bibinfo {author} {\bibfnamefont {C.}~\bibnamefont {Lester}},
  \bibinfo {author} {\bibfnamefont {M.~S.}\ \bibnamefont {Senn}}, \bibinfo
  {author} {\bibfnamefont {A.}~\bibnamefont {Bombardi}}, \ and\ \bibinfo
  {author} {\bibfnamefont {S.~M.}\ \bibnamefont {Hayden}},\ }\href@noop {}
  {\bibfield  {journal} {\bibinfo  {journal} {Phys. Rev. B}\ }\textbf {\bibinfo
  {volume} {89}},\ \bibinfo {pages} {224513} (\bibinfo {year}
  {2014})}\BibitemShut {NoStop}%
\bibitem [{\citenamefont {Thampy}\ \emph {et~al.}(2014)\citenamefont {Thampy},
  \citenamefont {Dean}, \citenamefont {Christensen}, \citenamefont {Steinke},
  \citenamefont {Islam}, \citenamefont {Oda}, \citenamefont {Ido},
  \citenamefont {Momono}, \citenamefont {Wilkins},\ and\ \citenamefont
  {Hill}}]{tham14}%
  \BibitemOpen
  \bibfield  {author} {\bibinfo {author} {\bibfnamefont {V.}~\bibnamefont
  {Thampy}}, \bibinfo {author} {\bibfnamefont {M.~P.~M.}\ \bibnamefont {Dean}},
  \bibinfo {author} {\bibfnamefont {N.~B.}\ \bibnamefont {Christensen}},
  \bibinfo {author} {\bibfnamefont {L.}~\bibnamefont {Steinke}}, \bibinfo
  {author} {\bibfnamefont {Z.}~\bibnamefont {Islam}}, \bibinfo {author}
  {\bibfnamefont {M.}~\bibnamefont {Oda}}, \bibinfo {author} {\bibfnamefont
  {M.}~\bibnamefont {Ido}}, \bibinfo {author} {\bibfnamefont {N.}~\bibnamefont
  {Momono}}, \bibinfo {author} {\bibfnamefont {S.~B.}\ \bibnamefont {Wilkins}},
  \ and\ \bibinfo {author} {\bibfnamefont {J.~P.}\ \bibnamefont {Hill}},\
  }\href@noop {} {\bibfield  {journal} {\bibinfo  {journal} {Phys. Rev. B}\
  }\textbf {\bibinfo {volume} {90}},\ \bibinfo {pages} {100510} (\bibinfo
  {year} {2014})}\BibitemShut {NoStop}%
\bibitem [{\citenamefont {Hunt}\ \emph {et~al.}(1999)\citenamefont {Hunt},
  \citenamefont {Singer}, \citenamefont {Thurber},\ and\ \citenamefont
  {Imai}}]{hunt99}%
  \BibitemOpen
  \bibfield  {author} {\bibinfo {author} {\bibfnamefont {A.~W.}\ \bibnamefont
  {Hunt}}, \bibinfo {author} {\bibfnamefont {P.~M.}\ \bibnamefont {Singer}},
  \bibinfo {author} {\bibfnamefont {K.~R.}\ \bibnamefont {Thurber}}, \ and\
  \bibinfo {author} {\bibfnamefont {T.}~\bibnamefont {Imai}},\ }\href@noop {}
  {\bibfield  {journal} {\bibinfo  {journal} {Phys. Rev. Lett.}\ }\textbf
  {\bibinfo {volume} {82}},\ \bibinfo {pages} {4300} (\bibinfo {year}
  {1999})}\BibitemShut {NoStop}%
\bibitem [{\citenamefont {Horibe}\ \emph {et~al.}(1997)\citenamefont {Horibe},
  \citenamefont {Inoue},\ and\ \citenamefont {Koyama}}]{hori97}%
  \BibitemOpen
  \bibfield  {author} {\bibinfo {author} {\bibfnamefont {Y.}~\bibnamefont
  {Horibe}}, \bibinfo {author} {\bibfnamefont {Y.}~\bibnamefont {Inoue}}, \
  and\ \bibinfo {author} {\bibfnamefont {Y.}~\bibnamefont {Koyama}},\
  }\href@noop {} {\bibfield  {journal} {\bibinfo  {journal} {Physica C}\
  }\textbf {\bibinfo {volume} {282--287}},\ \bibinfo {pages} {1071} (\bibinfo
  {year} {1997})}\BibitemShut {NoStop}%
\bibitem [{\citenamefont {Koyama}\ \emph {et~al.}(1995)\citenamefont {Koyama},
  \citenamefont {Wakabayashi}, \citenamefont {Ito},\ and\ \citenamefont
  {Inoue}}]{koya95}%
  \BibitemOpen
  \bibfield  {author} {\bibinfo {author} {\bibfnamefont {Y.}~\bibnamefont
  {Koyama}}, \bibinfo {author} {\bibfnamefont {Y.}~\bibnamefont {Wakabayashi}},
  \bibinfo {author} {\bibfnamefont {K.}~\bibnamefont {Ito}}, \ and\ \bibinfo
  {author} {\bibfnamefont {Y.}~\bibnamefont {Inoue}},\ }\href@noop {}
  {\bibfield  {journal} {\bibinfo  {journal} {Phys. Rev. B}\ }\textbf {\bibinfo
  {volume} {51}},\ \bibinfo {pages} {9045} (\bibinfo {year}
  {1995})}\BibitemShut {NoStop}%
\end{thebibliography}
\end{document}